\documentclass{aa}

\usepackage{breqn}
\usepackage{graphicx}
\usepackage{float}
\usepackage{placeins}
\usepackage{hyperref}
\usepackage{txfonts}

\usepackage{color}

\begin{document}

   \title{A high-resolution survey of protoplanetary disks in Lupus and the nature of compact disks}

   \author{Osmar M. Guerra-Alvarado,
           \inst{1}
           Nienke van der Marel,
           \inst{1}
           Jonathan P. Williams,
           \inst{2}
           Paola Pinilla,
           \inst{3}
           Gijs D. Mulders,
           \inst{4}
           Michiel Lambrechts,
           \inst{5}
           Mariana Sanchez
           \inst{1}
}
   \institute{Leiden Observatory, Leiden University, PO Box 9513, 2300 RA Leiden, The Netherlands.
              \email{guerra@strw.leidenuniv.nl}
    \and Institute for Astronomy, University of Hawai’i at Manoa, 2680 Woodlawn Dr., Honolulu, HI, USA
    \and  Mullard Space Science Laboratory, University College London,
3 Holmbury St Mary, Dorking, Surrey RH5 6NT, UK.
    \and Instituto de Astrof\'isica, Pontificia Universidad Cat\'olica de Chile, Av. Vicu\~na Mackenna 4860, 7820436 Macul, Santiago, Chile
    \and Center for Star and Planet Formation, GLOBE Institute, University of Copenhagen, Øster Voldgade 5–7, 1350 Copenhagen, Den- mark
             }

   \date{}

 
  \abstract
   {}
   {Most of the exoplanets discovered in our galaxy to date orbit low-mass stars, which tend to host small disks in their early stages. To better elucidate the link between planet formation and disk substructures, observational biases should be reduced through observations of these small, faint disks at the highest resolution using the Atacama Large Millimeter Array (ALMA).}
  {In this work, we present new high resolution (0.03-0.04'') ALMA observations at 1.3 mm of 33 disks located in the Lupus star-forming region that have total dust continuum fluxes < 25 mJy. Combining archival data and previously published work, we provide a near-complete high resolution image library of 73 protoplanetary (Class II) disks in the Lupus. This enable us to measure dust disk radii down to a limit of 0.6 au and analyze intensity profiles using visibility modeling.} 
  {We show that 67\% of Lupus protoplanetary dust disks have dust radii smaller than 30 au and characterize the newly discovered substructures in 11 disks with some of the shortest separation gaps. The size-luminosity relation, when accounting for the smallest disks, aligns well with a drift-dominated dust evolution scenario for the Class II Lupus disks.
  For the most compact disks, with radii less than 30 au, we compared measured sizes and fluxes to a grid of radiative transfer models to derive millimeter emitting dust masses, which ranged from 0.3 to 26.3 M$_{\oplus}$. Assuming that the detected substructures were dynamical effects of planets, we approximate the results of an interpolation to estimate planet masses and found a range of $20-2000\,$M$_\oplus$ with separations between 2 to 74 au. } 
  {Our results indicate that two-thirds of the protoplanetary disks in Lupus are smooth, on scales larger than 4 au, and compact, with substructures being more prominent in the few larger disks. These compact disks are consistent with drift-dominated evolution, with their masses and optical depths suggesting that they may have already experienced some planet formation, with most of the small solids converted into planetesimals and planets. This makes them prime candidates, with the optimal conditions, for explaining the formation and origin of super-Earths.} 

   \keywords{Planetary systems: Protoplanetary disk - Radio continuum: planetary systems
               }
\titlerunning{High-resolution view of Lupus disks}
\authorrunning{O. M. Guerra-Alvarado et al.}
   \maketitle
%

\section{Introduction}
The field of protoplanetary disks has been revolutionized in the past decade with the sensitivity and spatial resolution of the ALMA telescope. Significant efforts have been dedicated to studying the demographics of protoplanetary disks in recent years with ALMA, providing a broader perspective on disk evolution and enabling direct comparisons and statistical analyses of exoplanets in conjunction with protoplanetary disks (\citet{2018ApJ...869L..47Z}; \citet{2019MNRAS.486..453L}; \citet{2021AJ....162...28V}).
Early ALMA survey programs yielded valuable insights on dust evolution in protoplanetary disks through various disk relations \citep[e.g.,][]{Pinilla2020}, even though they were taken at relatively low resolution ($\sim$0.25", \citet{2016ApJ...828...46A}; \citet{2016ApJ...827..142B}; \citet{2019MNRAS.482..698C}). Subsequent high-resolution  observations ($\sim 0\farcs04$) focused on the brightest disks, which are often large and harbor numerous extended substructures (\citet[]{2018ApJ...869L..41A}; \citet{2021MNRAS.501.2934C}), usually associated with pressure bumps halting the radial drift through dust traps \citep{Pinilla2012}. However, many disks remained unresolved in this initial reconnaissance, preventing a thorough investigation of the entire disk population. 

The first discovery of a very small dust disk with high-resolution ALMA observations was XZ Tau B, with a dust disk size of only 3.4 au and potentially an inner cavity, located within a binary system of 39 au separation \citep{Osorio2016,Ichikawa2021}, consistent with predictions from radial drift in binary systems \citep{Zagaria2021}. The first small disk in a single star system studied at high resolution was CX Tau, analyzed by \citet{2019A&A...626L...2F}. They found no substructure but measured a dust disk radius of 14 au, which was five times smaller than the CO extent, indicating efficient radial drift. In subsequent years, several additional studies (\citet{2019ApJ...882...49L}; \citet{2021A&A...645A.139K}; \citet{2022arXiv220408225V}; \citet{2024A&A...682A..55M}; \citet{2024ApJ...966...59S}) have identified an increasing number of compact disks, defined as having dust disk radii smaller than 30 au for the purposes of this paper, which is also the size of Neptune's orbit.
We have identified 33 such compact disks in the literature to date, a few of which show centrally cleared cavities. However, the size distribution and structural classification of disks within a single star-forming region remain limited by the need for larger sample sizes of high-resolution observations and more accurate measurements of the disk radii for the smallest disks that often remain unresolved.

One of the most important relationships that remains incomplete for the most compact disks and within single star-forming regions is the Size-Luminosity Relation (SLR). Assuming that the relation is not primordial, meaning that low-mass disks are born small while high-mass disks are born large, current literature identifies three distinct slopes that define the behavior and evolution of disks within this relation. Dust grain growth in disks is limited by two main barriers \citep{2012A&A...539A.148B}: the drift barrier, largely driven by radial drift \citep{1977MNRAS.180...57W}, and the fragmentation barrier, shaped by turbulence which causes grain collisions and destruction \citep{1980A&A....85..316V}. These barriers determine how disks evolve along the SLR.
\citet{2019MNRAS.486L..63R} studied the SLR in the context of these two primary dust growth barriers, introducing two distinct slopes in this relation. The first is the drift-dominated slope, where a disk’s position along the SLR is mainly influenced by its dust mass, affecting both its luminosity and radius. For disks with a smooth density profile, this slope follows $F_{mm} \propto R_{eff}^{2}$ for its flux $F_{mm}$ and disk size $R_{eff}$, respectively, where $R_{eff}$ is typically defined as the radius enclosing a specified fraction (often 68\% or 90\%) of the total flux from the disk. However, in fragmentation-dominated disks, the slope becomes steeper, following $F_{mm} \propto R_{eff}$, as fragmentation increases the mass retained in the disk's, enhancing the flux for the same radius compared to the drift-dominated scenario, thereby altering the disk's position on the SLR.  
The third slope, described by \citet{2022A&A...661A..66Z}, is the trap-dominated slope, which implies that disks have strong dust traps. In this case, the SLR behaves differently compared to smooth, non-trap disks. This slope assumes that a disk begins its evolution with a planet forming within it, shifting its evolution along the SLR. The relation for these disks follows $F_{mm} \propto R_{eff}^{5/4}$, which is steeper than the drift-dominated slope but less steep than the fragmentation-dominated one. More recently, \citet{2024A&A...688A..81D} continued this exploration, suggesting that even smooth disks might have hidden substructures. They were able to reproduce the observed $F_{mm} \propto R_{eff}^{2}$ slope by having optically thick high-flux and optically thin low-flux disks with substructures.

Several factors contribute to deviations from the SLR, as noted by \citet{2022A&A...661A..66Z}. Dust properties, such as variations in opacity or porosity, can shift a disk's position along the SLR by affecting its luminosity without changing its size. Additionally, the turbulence parameter $\alpha$ plays a significant role in modifying a disk's location on the SLR. Both factors are essential for the drift and fragmentation barriers, which in turn are key to determining dust evolution and growth. This highlights the importance of studying the SLR in greater detail and at the highest possible resolution, as recent research has done. 

Furthermore, constraining disk parameters and extrapolating the SLR to the smallest disks results in disk sizes of only a few au which has significant implications for understanding dust substructures and the comparison with exoplanet populations. Recently, there have been several efforts to directly link exoplanet observations to the properties of their birthplace, the protoplanetary disks. Key parameters have been studied, such as the available bulk mass in protoplanetary disks to form giant and terrestrial exoplanets (\citet{2018A&A...618L...3M}; \citet{2021ApJ...920...66M}), as well as the connection between disk substructures and exoplanet demographics (\citet{2019MNRAS.486..453L}; \citet{2021AJ....162...28V}; \citet{2023ApJ...952..108Z}). Specifically, \citet{2021AJ....162...28V} aimed to understand whether disk substructures influence exoplanet formation scenarios or are linked to the observed disk dichotomy. Their work suggested that the majority of close-in rocky exoplanets around M-stars have likely formed in the more abundant smooth, compact disks in the absence of giant planets at wide orbits which would prevent radial drift \citep{2021ApJ...920L...1M}. These compact disks can form planets, particularly super-Earths through pebble accretion, under the influence of substantial radial drift, concentrating sufficient material in the inner regions \citep{2024A&A...689A.236S}, under the assumption that the bulk of the initial dust mass in embedded disks (Class 0/I) decreases rapidly to the mass in the protoplanetary disk stage (Class II) through radial drift (Appelgren et al.\ subm.). However, due to the relatively low resolution of observations of protoplanetary disks, compared to the regions where most exoplanets are found, the relation between such disks and rocky exoplanets, if any, remains inconclusive.
\begin{table*}[!h]
\caption{ALMA images characteristics of the new observations from projects 2022.1.00154.S and 2018.1.01458.S}

\centering

\resizebox{0.95\textwidth}{!}{%
\begin{tabular}{cccccccc}
\label{tab:data_reduction}

Source &2MASS& F$_{1.3mm}$ & Weighting & Beam size & Rms & Peak SNR& Visible \\
&Identifier&[mJy]&&["]&[mJy $\cdot$ beam$^{-1}$]& &Substructure\\
\hline
\hline
J16124373-3815031 &J16124373-3815031& 11.54 $\pm$ 0.04 & Briggs (2.0) & 0.048$\times$0.032 & 0.04 & 16.32& No \\
Sz117 &J16094434-3913301& 3.79 $\pm$ 0.05 & Briggs (0.5) & 0.036$\times$0.030 & 0.045 & 14.4& No \\
Sz110 &J16085157-3903177& 6.59 $\pm$ 0.04 & Briggs (1.0) & 0.047$\times$0.038 & 0.039 & 18.56& No \\
J16134410-3736462 &J16134410-3736462&0.56 $\pm$ 0.04 & Briggs (2.0) & 0.048$\times$0.032 & 0.039 & 11.3& No \\
J16080017-3902595 &J16080017-3902595& 1.14 $\pm$ 0.04 & Natural & 0.051$\times$0.044 & 0.039 & 15.47 & No\\
Sz69 &J15451741-3418283& 8.36 $\pm$ 0.12 & Briggs (0.0) & 0.049$\times$0.032 & 0.128 & 10.46& No \\
Sz95 &J16075230-3858059& 1.63 $\pm$ 0.04 & Briggs (1.0) & 0.046$\times$0.038 & 0.038 & 19.0& No \\
J16085373-3914367 &J16085373-3914367& 1.4 $\pm$ 0.03 & Briggs (0.5) & 0.041$\times$0.032 & 0.04 & 7.74& No \\
Sz88A &J16070061-3902194& 3.11 $\pm$ 0.06 & Briggs (0.0) & 0.029$\times$0.024 & 0.058 & 22.8& No \\
J16073773-3921388 &J16073773-3921388& 0.73 $\pm$ 0.04 & Natural & 0.051$\times$0.044 & 0.037 & 7.58& No \\
J16002612-4153553 &J16002612-4153553& 0.61 $\pm$ 0.03 & Briggs (1.0) & 0.032$\times$0.024 & 0.029 & 14.35& No \\
Sz102 &J16082972-3903110& 5.12 $\pm$ 0.04 & Briggs (0.5) & 0.036$\times$0.030 & 0.043 & 18.8& No \\
Sz113 &J16085780-3902227& 9.52 $\pm$ 0.05 & Briggs (0.5) & 0.036$\times$0.030 & 0.04 & 20.13& No \\
Sz97 &J16082180-3904214& 1.74 $\pm$ 0.05 & Briggs (0.5) & 0.036$\times$0.030 & 0.045 & 15.09 & No\\
J16085324-3914401 &J16085324-3914401& 7.56 $\pm$ 0.04 & Briggs (0.5) & 0.040$\times$0.032 & 0.042 & 21.74& No \\
Sz77 &J15514695-3556440& 1.49 $\pm$ 0.04 & Briggs (0.5) & 0.034$\times$0.030 & 0.04 & 16.66& No \\
Sz130 &J16003103-4143369& 1.07 $\pm$ 0.03 & Briggs (1.0) & 0.032$\times$0.024 & 0.031 & 26.66& No \\
Sz106 &J16083976-3906253& 0.75 $\pm$ 0.04 & Briggs (1.0) & 0.046$\times$0.038 & 0.039 & 10.35& No \\
V1192Sco &J16085143-3905304& 0.32 $\pm$ 0.02 & Natural & 0.04$\times$0.044 & 0.038 & 5.74& No \\
Sz81A &J15555030-3801329& 4.0 $\pm$ 0.04 & Briggs (0.5) & 0.034$\times$0.030 & 0.042 & 26.9& No \\
Sz81B &J15555030-3801329& 1.26 $\pm$ 0.04 & Briggs (0.5) & 0.034$\times$0.030 & 0.042 & 14.56 & No\\
Sz74A &J15480523-3515526& 7.97 $\pm$ 0.05 & Briggs (0.0) & 0.027$\times$0.023 & 0.054 & 30.0& No \\
Sz74B &J15480523-3515526& 3.3 $\pm$ 0.05 & Briggs (0.0) & 0.027$\times$0.023 & 0.054 & 42.43& Yes \\
V856ScoB &J16083427-3906181& 7.19 $\pm$ 0.11 & Uniform  & 0.029$\times$0.023 & 0.11 & 17.74& No \\
V856ScoA &J16083427-3906181& 21.74 $\pm$ 0.11 & Uniform & 0.029$\times$0.023 & 0.11 & 44.13& No \\
J15450887-3417333 &J15450887-3417333& 22.27 $\pm$ 0.10 & Briggs (0.5) & 0.063$\times$0.038 & 0.091 & 19.28& No \\
J16075475-3915446 &J16075475-3915446& 0.43 $\pm$ 0.04 & Natural & 0.051$\times$0.044 & 0.038 & 6.9& No \\
J16084940-3905393 &J16084940-3905393& 0.51 $\pm$ 0.04 & Briggs (0.5) & 0.036$\times$0.030 & 0.041 & 11.97& No \\
J15592523-4235066 &J15592523-4235066& 0.31 $\pm$ 0.06 & Briggs (2.0) & 0.047$\times$0.039 & 0.035 & 7.34 & No\\
Sz108B &J160842.9-390615& 10.38 $\pm$ 0.04 & Briggs (1.0) & 0.046$\times$0.039 & 0.039 & 17.05& Yes (Cavity) \\
J16092697-3836269 &J16092697-3836269& 1.72 $\pm$ 0.04 & Briggs (1.0) & 0.47$\times$0.038 & 0.04 & 7.35& Yes (Cavity) \\
Sz72 &J15475062-3528353& 5.40 $\pm$ 0.03 & Briggs (1.0) & 0.039$\times$0.034 & 0.039 & 18.93& Yes (Cavity)\\
Sz90 &J16071007-3911033& 6.9 $\pm$ 0.04 & Briggs (1.0) & 0.047$\times$0.038 & 0.04 & 15.425& Yes (Cavity) \\
Sz96 &J16081263-3908334& 1.41 $\pm$ 0.04 & Briggs (0.5) & 0.048$\times$0.030 & 0.045 & 8.12& Yes (Cavity) \\
Sz123A &J16105158-3853137& 18.11 $\pm$ 0.07 & Briggs (1.0) & 0.071$\times$0.063 & 0.071 & 16.74& Yes (Cavity)\\
Sz100 &J16082576-3906011& 24.86 $\pm$ 0.07 & Briggs (1.0) & 0.071$\times$0.063 & 0.067 & 17.5& Yes (Cavity)\\
Sz131 &J16004943-4130038& 3.44 $\pm$ 0.03 & Briggs (1.5) & 0.035$\times$0.026 & 0.03 & 11.08& Yes (Cavity)\\
Sz73 &J15475693-3514346& 3.67 $\pm$ 0.06 & Briggs (0.0) & 0.048$\times$0.033 & 0.078 & 35.63& Yes (Gap) \\
Sz66 &J15392828-3446180& 6.06 $\pm$ 0.04 & Briggs (0.5) & 0.033$\times$0.027 & 0.043 & 17.85& No \\
Sz65 &J15392776-3446171& 21.08 $\pm$ 0.04 & Briggs (0.5) & 0.033$\times$0.027 & 0.043 & 16.6 & No\\
J16083070-3828268 &J16083070-3828268& 43.96 $\pm$ 0.07 & Briggs (2.0) & 0.079$\times$0.072 & 0.07 & 25.81& Yes (Cavity) \\
Sz98&J16082249-3904464& 119.21 $\pm$ 0.06 & Briggs (1.0) & 0.071$\times$0.063 & 0.07 & 18.89& Yes (Gaps)\\
\hline
\end{tabular}%
}
\end{table*}
\begin{figure*}[!h]

\centering
\includegraphics[width=2\columnwidth,trim={0cm 0.0cm 0cm 0.0cm}]{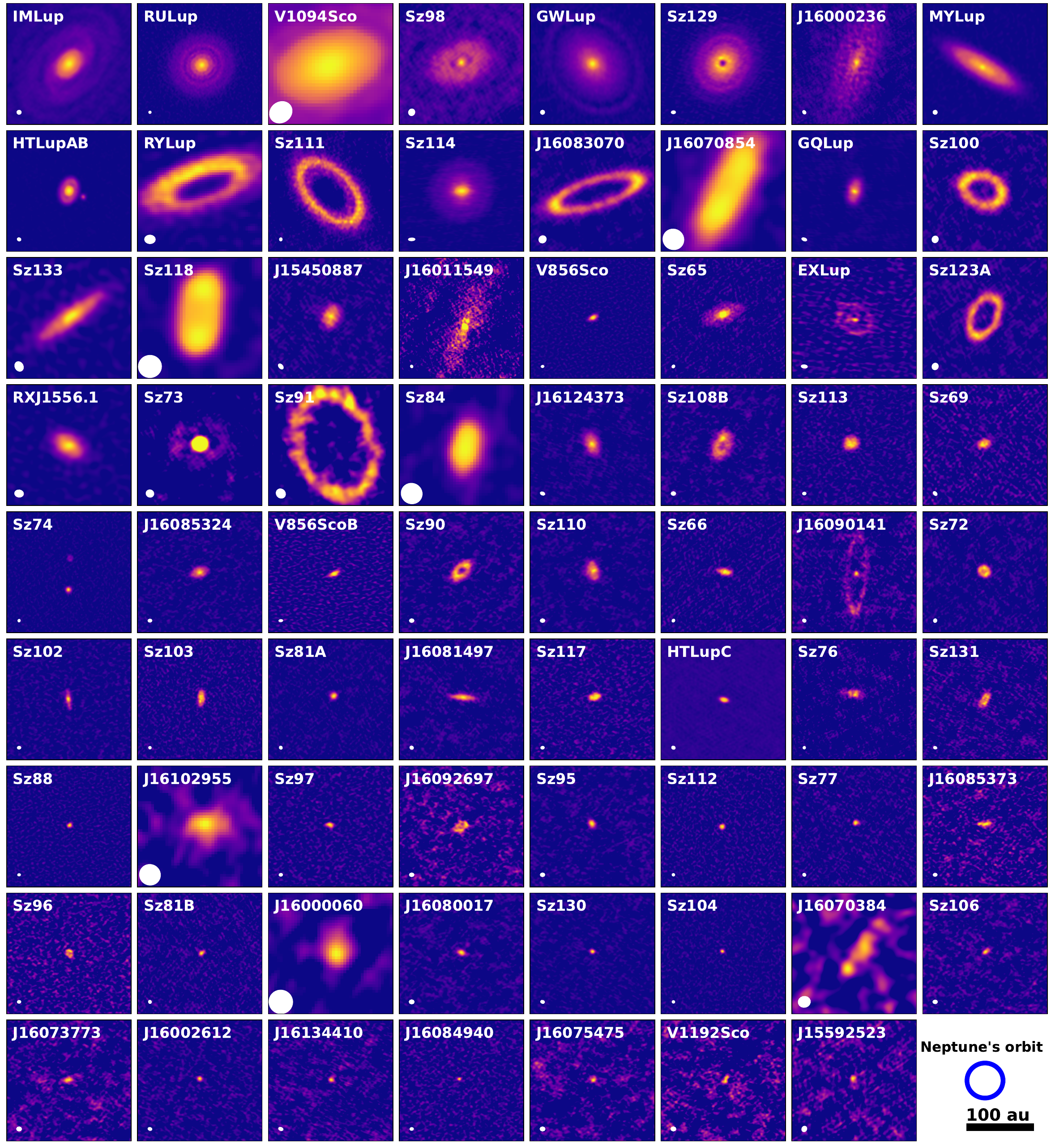}
\caption{\em{All protoplanetary disks in Lupus at high resolution from various projects and observations with ALMA (see Section 2), shown at the same spatial scale. The disks are arranged in descending order based on their total flux. The scale bar and Neptune's orbit in the final panel assumes that each disk is at a distance of 160 pc.} Several cavities and substructures are observed, though the smallest disks are barely visible.}
\label{fig:disk_images}
\end{figure*}
\begin{figure*}[!h]
\label{fig:compact_disks}
\centering
\includegraphics[width=2\columnwidth,trim={0cm 0.0cm 0cm 0.0cm}]{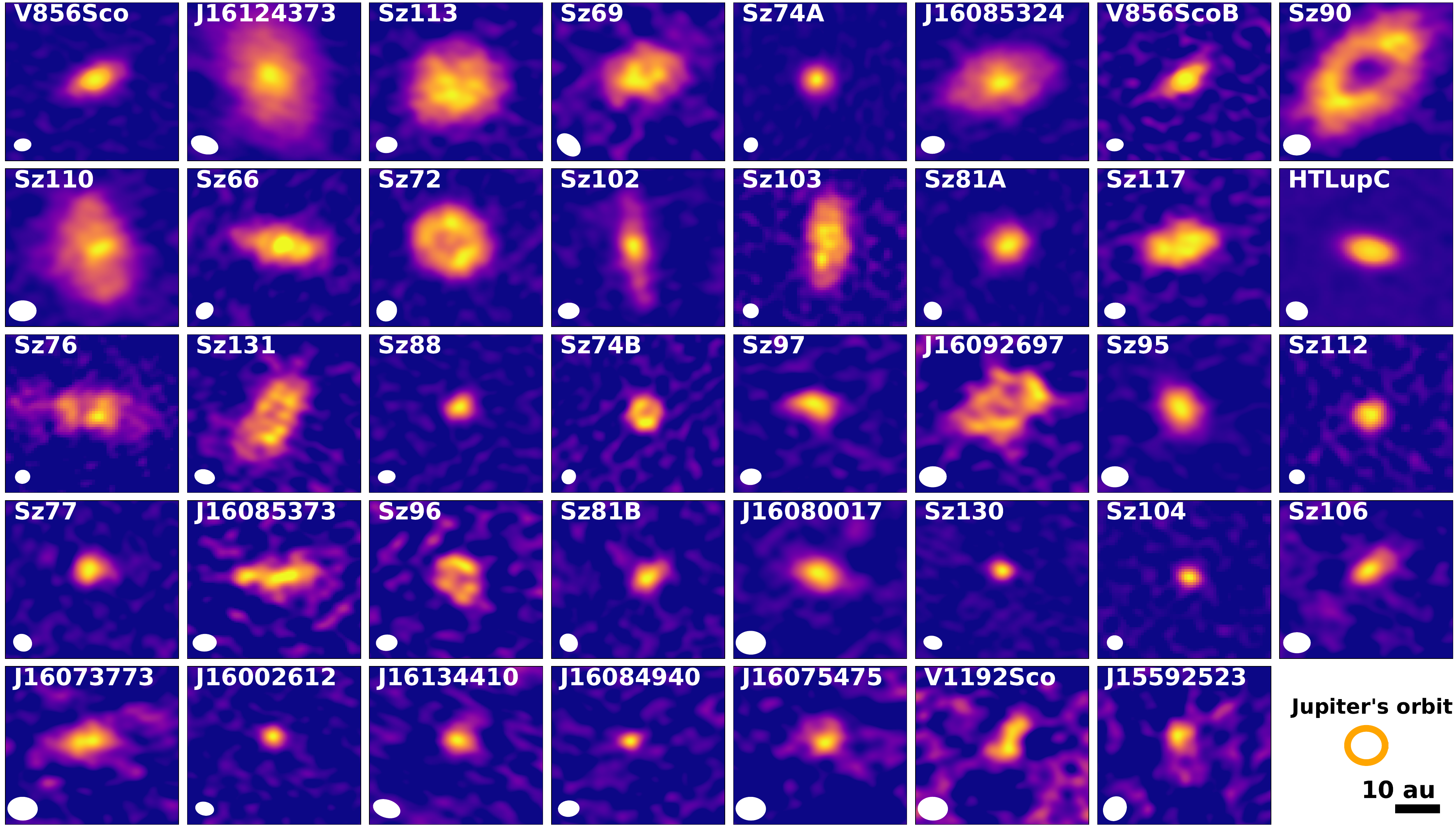}
\caption{\em{Zoom-in on the smallest disks in the Lupus molecular cloud. Some of these small disks exhibit substructure, although most appear featureless. The disks are arranged in descending order of their total flux, consistent with the larger disks shown previously. The scale bar and Jupiter's orbit in the final panel assumes that each disk is at a distance of 160 pc.}
}
\end{figure*}

In this work, we present a near-complete high-resolution survey of 73 Class II protoplanetary disks in the young, nearby Lupus star-forming region. The Lupus region, part of the Scorpius-Centaurus association, consists of several subgroups (Lupus 1–9), each associated with distinct molecular clouds. It is a young (1–2 Myr) and nearby (150–200 pc) region, comparable in proximity and age to Taurus \citep{2008hsf2.book..295C}.
This study is based on new ALMA continuum observations as well as ALMA archival data. We measure dust disk radii down to 0.6 au, fit radial profiles to determine substructures and study the relation between the disk flux density and its size.
We discuss the observations and data reduction in \S\ref{sec:obs}, the analysis of the continuum visibilities for measuring disk sizes and substructure in \S\ref{sec:results}, and discuss the implication of our results and comparison with a model grid in \S\ref{sec:discussion}. We then summarize our findings in \S\ref{sec:summary}.
\section{Observations}
\label{sec:obs}
Our data were obtained from Cycle 9 observations with the Atacama Large Millimeter/submillimeter Array (ALMA) at Band 6. These observations, with project code 2022.1.00154.S and PI Nienke van der Marel, were carried out in 10 execution blocks between August 16, 2023, and September 11, 2023. The total observing time was 8.89 hours, with $\sim$ 7.25 minutes spent on each source. In total, 33 sources were observed: Sz65, Sz66, Sz77, Sz72, Sz74A, Sz74B, J15592523-4235066, J16002612-4153553, Sz130, Sz131, Sz81A, J16073773-3921388, J16075475-3915446, J16080017-3902595, J16084940-3905393, J16085324-3914401, J16085373-3914367, J16092697-3836269, Sz102, Sz106, Sz108B, Sz110, Sz113, Sz117, Sz88A, Sz88B, Sz90, Sz95, Sz96, Sz97, V1192 Sco, V856 Sco A, and V856 Sco B. The final datasets for each source include 6 spectral windows with frequencies between 219–234 GHz and a total bandwidth of 2 GHz each for the continuum, as well as 6 spectral windows between 220–231 GHz with a total bandwidth of 1.875 GHz dedicated to $^{12}$CO and $^{13}$CO line emission, all with a maximum recoverable scale between 0.37" and 0.715". Additionally, we incorporated archival ALMA data to complete the sample of Lupus disks at high resolution (project code: 2018.1.01458.S, PI: Yen, Hsi-Wei). These observations were conducted between July 18, 2019, and July 19, 2019, with a total integration time of 1.1 hours and a total of 8 sources: Sz71, Sz69, J15450887-3417333, Sz98, Sz123A, Sz100, J16083070-3828268, and Sz73. These observations include 6 spectral windows to observe line emission, each with a bandwidth of 58.6 MHz, and only one spectral window for the continuum with a total bandwidth of 2 GHz, between 232-234 GHz. Finally, a Band 4 dataset was used for EX Lup, as no high-resolution observations were available at any other band (project code: 2017.1.00388.S, PI: Liu, Hauyu Baobab). These observations were conducted on November 11, 2017, covering a frequency range of 145 to 161 GHz, with a total integration time of 5.1 minutes. We searched for line emission to flag but found none, indicating only the presence of the continuum. However, this data were not used for any part of the analysis and served only to provide a high-resolution image of the source.

All datasets were calibrated using the pipeline and scripts provided by the ALMA staff. Version 6.5.4.9 of the Common Astronomy Software Applications (CASA; \citealt{2007ASPC..376..127M}) was utilized to analyze and process the data, as well as to clean and create the final images. We separated the line and continuum emission spectral windows for each source and averaged the channels in the continuum spectral windows. Self-calibration was attempted on the new datasets, but due to the limited observing time per source and the low signal-to-noise ratios (SNR), it was not possible to achieve satisfactory results. We used the \textsc{tclean} task in CASA with the MTMFS deconvolver \citep{2011A&A...532A..71R}, employing scales of 0, 1, 3, and 5 times the beam size to create the final images of each disk. Several weightings were tested depending on the source, and different weightings were selected from the datasets, as shown in Table~\ref{tab:data_reduction}. Natural weighting emphasizes short baselines, improving sensitivity but at the cost of resolution. In contrast, uniform weighting gives more weight to longer baselines, maximizing resolution but reducing sensitivity and making the images noisier while Briggs weighting serves as a compromise between natural and uniform, balancing both sensitivity and resolution. For each source, we selected the weighting that offered the highest resolution while maintaining sufficient sensitivity to capture and distinguish the entire disk structure effectively. For all sources, pixel sizes between 0.001 and 0.003 arcseconds were used to ensure the pixel size was approximately ten times smaller than the beam size for all images.

To complete the high-resolution Lupus disk sample, we supplemented our data with images obtained from several previous studies. These included Sz103, Sz76, Sz104, Sz112, J16011549-4152351, J16081497-3857145, J16000236-4222145, J16090141-3925119, and J16070384-3911113 from \citep{2022arXiv220408225V}; HT Lup A, B, and C, GW Lup, IM Lup, RU Lup, Sz114, Sz129, and MY Lup from the Disk Substructures at High Angular Resolution Project (DSHARP; \citealt{2018ApJ...869L..41A}); GQ Lup from \citealt{2017ApJ...836..223W}; and Sz91 in Band 4 from \citealt{2021ApJ...923..128M}. Additionally, we used RY Lup data from \citealt{2020ApJ...892..111F}. For already resolved disks J16070854-3914075, Sz118, Sz84, as well as for two sources without high-resolution observations (J16102955-3922144 and J16000060-4221567), we used images from \citealt{2018ApJ...859...21A}. For V1094 Sco, we used the image presented in \citealt{2018A&A...616A..88V}, and for Sz111, data from (Rota et al. in prep) (project code 2018.1.00689.S, PI: Muto, Takayuki). Finally, RXJ1556.1-3655 and Sz133 data were sourced from Bosschaart et al. in prep (project code: 2022.1.01302.S, PI: Mulders, Gijs). The total sample consists of 73 disks in Lupus with all but 5 disks observed at very high angular resolution of $\lesssim$0.05". An overview of the characteristics of the images created in this paper is presented in Table~\ref{tab:data_reduction}, with all the images of the disks shown in Figure~\ref{fig:disk_images}. The most compact disks with radii < 0.15" are displayed on a smaller scale in Figure~\ref{fig:compact_disks}.

\section{Results}
\label{sec:results}
\subsection{Dust continuum images}

In this sample of Class II Lupus disks with continuum images at high angular resolution we find a range of morphologies across all sizes. Although most substructures are detected in the more massive and larger disks, several compact disks also exhibit substructure, such as gaps and inner cavities. Given our resolution, which limits detection to structures as small as approximately 4 au ($\sim$ 0.03 arcseconds), we may still miss small inner cavities. With this in mind, the smallest cavity detected in the image plane is in Sz72, with a radius of 4.1 au, right at the edge of our resolution limit. 
Most disks are well resolved, however, a few remained only marginally resolved. Notably, Sz104, J16075475-3915446, J15592523-4235066, and J16084940-3905393 are very poorly resolved even at $0\farcs 03$ resolution, which implies a radii of less than 2 au. J16080017-3902595 and J16000060-4221567 are also unresolved at $0\farcs 25$ but lack the high resolution data of the bulk of the sample.

In total, we resolved 11 new disks with substructures: 10 cavities (J16083070, Sz100, Sz123A, EXLup, Sz108B, Sz90, Sz72, Sz131, J16092697, and Sz96) and 1 very faint ring and gap (Sz73). We have a total of 6 resolved binary systems with disk detections and within the field of view of the observations; Sz66 - Sz65, Sz74A - Sz74B, V56ScoA - V856ScoB, Sz81A - Sz81B, J16085324 - 3914401-J16085373-3914367 and the HT-Lup triple system. However, no disk detections were found in three additional well-known binaries: Sz88B, Sz108A, and Sz123B-C \citep{2021MNRAS.501.2305Z}. The total flux for all targets is determined through aperture photometry, where the aperture is extended until the flux flattens, and inclination corrections are applied. The uncertainty is calculated as the standard deviation inside each aperture. Most flux measurements fall within an 10$\%$ uncertainty range compared to those reported at low resolution by \citealt{2018ApJ...859...21A}. Some discrepancies could arise from the absence of short baselines in our observations, potentially leading to the loss of extended emission flux. Overall, the fluxes appear consistent with previous results, meaning there is no need for a detailed re-evaluation and we proceed to the analysis.

\subsection{Visibility modeling}

To analyze the exact orientation and morphologies of these disks, we used the \textsc{Galario} code \citep{2018MNRAS.476.4527T} to model the visibilities of the observations with the best fit possible. We did not fit any features that were not already visible in the continuum images. For all isolated disks, we computed the visibilities from an axisymmetric brightness profile using different models. For each model, we fitted the visibilities, $V_{mod}$, following the approach outlined in \citealt{2018MNRAS.476.4527T}. We explored the parameter space of each model using a Bayesian approach, employing the Markov chain Monte Carlo (MCMC) ensemble sampler provided by \textsc{EMCEE} \citep{2013PASP..125..306F}. Each model generates a brightness profile, which is then transformed and compared to the observed visibilities. Posterior distributions are obtained by assuming a Gaussian likelihood. In this paper, we employ three different models for the isolated disks choosing the most simple structure we observed for each disk in the continuum images;\\ 

Gaussian like disk:
\begin{equation}
I(R) = I_0 \cdot \exp\left(-\frac{1}{2} \cdot\left(\frac{R}{r_{c}}\right)^2\right)
\end{equation}
Where I$_0$ is the peak intensity, R is the radial distance from the center and $r_{c}$ is the width of the Gaussian.

A ring with different slopes on each side:
\begin{equation}
\begin{aligned}
I(R) = & \, I_{0} \left( \left(1 - \Theta(R -r_{\rm ring})\right) \exp\left(\frac{-(R - r_{\rm ring})^2}{2 (r_{w}a)^2}\right) \right) \\
& + I_{0} \left( \Theta(R - r_{\rm ring}) \exp\left(\frac{-(r - r_{\rm ring})^2}{2 (r_{w}b)^2}\right) \right)
\end{aligned}
\end{equation}
Where I$_0$ is the peak intensity of the ring, $r_{\rm ring}$ is the peak position of the ring, $r_{w}a$ the width of the ring on the inner side, $r_{w}b$ the width of the ring on the outer side and $\Theta$ is the step function:
\begin{equation}
\Theta(x) =
\begin{cases} 
1 & \text{if } x \geq 0 \\
0 & \text{if } x < 0
\end{cases}
\end{equation}
With the first term active when inside the position of the ring peak, and the second term when outside the position of the ring peak.

Finally, only one disk, Sz 73, required a two-component model to describe an inner Gaussian and a faint ring. We modeled this disk as follows:

\begin{equation}
\begin{aligned}
I(R) =& I_0 \cdot \exp\left( -\frac{1}{2} \cdot \left( \frac{R}{r_{c}} \right)^2 \right) \\
&+ I_0b \cdot \exp\left( -\frac{1}{2}\cdot \left( \frac{R - r_{\rm ring}}{r_{\rm width}} \right)^2 \right)
\end{aligned}
\end{equation}
Where the left Gaussian describes a ring with $I_0b$ as the intensity at the peak of the ring, $r_{\rm ring}$ as the radial position of the peak of the ring and $r_{\rm width}$ as the width of the ring.

For the binaries, we sampled a 2D image and computed the corresponding visibilities, rather than using a single radial profile. We adopted a method similar to that of \citet{2019A&A...628A..95M}, where the visibilities of both sources were summed and compared to a Gaussian likelihood to obtain the final visibilities. We used the previously described Gaussian model, where the total visibilities are expressed as the sum of the Fourier transforms of each of the Gaussian models. This is mathematically represented as follows
\begin{equation}
V_{modT}=V_{modGauss1}+V_{modGauss2}
\end{equation}
For two of the three binaries in this study, we employed a model consisting of two Gaussian profiles, which were added together in the visibility plane. The model, $V_{modGauss1}$ is a function of the brightness profile parameters and includes an offset relative to the phase center of the image.

V856ScoA was a special case where we see significant flux even on the longest baselines indicating an unresolved point source in addition to a Gaussian disk. To account for this, we defined a combined model as follows:
\begin{equation}
V_{\text{modGauss1}} = \mathcal{F} \left( I_{0} \cdot \exp\left(-\frac{1}{2}\cdot \left(\frac{R}{r_{c}}\right)^2 \right) \right) + \mathcal{F}\left( \delta(R) \right)
\end{equation}
Where $\delta(R)$ is defined as a Gaussian with an extremely small width, effectively approximating a delta function.
For all the single disks, we used \textsc{uvmodelfit} within CASA to obtain the initial parameter estimates. For the MCMC sampling, we employed between 48 and 120 walkers, running the chains for 2,000 to 30,000 steps to ensure convergence, with the binaries requiring more walkers and iterations to achieve satisfactory results. To verify convergence, we discarded the first 1,000 steps (burn-in phase) and examined the corner plots. The final model parameters are represented by the median of the posterior probability distributions for each parameter, while the uncertainties are given by the 16th and 84th percentiles.

Subsequently, we derived the total flux and dust disk radius from the modeled images. The results for the 38 sources modeled are presented in Table~\ref{tab:galariofit} with the corresponding distribution of disk radii shown in Figure~\ref{fig:size_distribution}. We excluded Sz65, Sz66, J16083070, and Sz98, from the \textsc{galario} analysis as these have been previously analyzed in \citet{2024A&A...682A..55M}, \citet{2019A&A...624A...7V}, and \citet{2023A&A...679A.117G}. The corresponding model images are shown in Figure~\ref{fig:galariomodels} and the visibility plots in Appendix \ref{appendix:b}.

\begin{table*}

\caption{Galario visibility fitting results of the disks in Table 1, excluding Sz66, Sz65, J16083070-3828268 and sz98. \textsuperscript{a} Due to the small sizes and low brightness, the PA and i are assumed zero for these targets. Therefore, R$_{68\%}$ and R$_{90\%}$ should be interpret with caution.}
 \setlength\tabcolsep{3.pt}
\renewcommand{\arraystretch}{1.35}
\centering
\resizebox{\textwidth}{!}{%
\begin{tabular}{lcccccccccccccc}

\hline
\multicolumn{12}{c}{Gaussian Models}\\
\hline

Source & log I$_{0}$ & r$_{c}$ &&&& Inc & P.A & dRa & dDec & F$_{1.3mm}$ & R$_{68}$&R$_{90}$ \\ 
 & [Jy sr$^{-1}$]& ["] & &&& [$^{\circ}$] &[$^{\circ}$]&["]&["]&[mJy]&[au]&[au]\\
\hline
\hline
J16124373-3815031 & $10.33^{+0.005}_{-0.005}$ & $0.07^{+0.0008}_{-0.0008}$ & &&&$51.92^{+0.72}_{-0.74}$ & $16.26^{+0.9}_{-0.95}$ & $0.26^{+0.0005}_{-0.0006}$ & $-0.02^{+0.0007}_{-0.0007}$ & 11.33 & 14.83&21.84 \\ 
Sz117& $10.43^{+0.01}_{-0.01}$ & $0.041^{+0.0009}_{-0.0009}$ & &&&$55.05^{+1.34}_{-1.47}$ & $105.22^{+1.52}_{-1.608}$ & $0.27^{+0.0007}_{-0.0007}$ & $-0.17^{+0.0005}_{-0.0005}$ & 3.99 & 7.7&11.56 \\ 
Sz110 & $10.22^{+0.007}_{-0.007}$ & $0.064^{+0.0008}_{-0.0008}$ & &&&$49.55^{+1.10}_{-1.15}$ & $13.10^{+1.45}_{-1.43}$ & $-0.002^{+0.0006}_{-0.0007}$ & $-0.083^{+0.0008}_{-0.0008}$ & 6.73 & 12.69&18.51 \\ 
J16134410-373646 & $10.53^{+0.10}_{-0.089}$ & $0.013^{+0.002}_{-0.002}$ & &&&$49.40^{+15.71}_{-32.5}$ & $155.7^{+13.42}_{-44.2}$ & $-0.08^{+0.001}_{-0.001}$ & $-0.14^{+0.001}_{-0.001}$ & 0.58 & 2.79&3.95\\ 
J16080017-3902595 & $10.44^{+0.04}_{-0.04}$ & $0.02^{+0.002}_{-0.002}$ & &&&$65^{+3.28}_{-3.8}$ & $64.54^{+3.93}_{-3.93}$ & $0.0019^{+0.001}_{-0.001}$ & $-0.09^{+0.0009}_{-0.0009}$ & 1.11 & 4.52& 7\\ 
Sz69 & $10.6^{+0.013}_{-0.012}$ & $0.049^{+0.001}_{-0.001}$ & &&&$54.56^{+1.46}_{-1.62}$ & $115.87^{+1.85}_{-1.93}$ & $-0.02^{+0.0007}_{-0.0008}$ & $-0.07^{+0.0006}_{-0.0006}$ & 8.19 & 8.98&13.37 \\ 
Sz95 & $10.52^{+0.034}_{-0.032}$ & $0.027^{+0.001}_{-0.001}$ & &&&$63.30^{+2.64}_{-2.87}$ & $21.14^{+2.77}_{-2.81}$ & $0.1^{+0.0008}_{-0.0007}$ & $-0.27^{+0.0009}_{-0.001}$ & 1.65 & 4.93& 7.56\\ 
J16085373-3914367 & $10.38^{+0.178}_{-0.11}$ & $0.048^{+0.002}_{-0.004}$ & &&&$81.83^{+2.67}_{-3.90}$ & $93.04^{+1.4}_{-1.47}$ & $0.057^{+0.002}_{-0.002}$ & $-0.17^{+0.0008}_{-0.0008}$ & 1.22 & 7.18 &12\\ 
Sz88A & $11.09^{+0.013}_{-0.014}$ & $0.014^{+0.0004}_{-0.0005}$ & &&&$35.30^{+3.0}_{-4.74}$ & $138.94^{+6.48}_{-8.27}$ & $-0.15^{+0.0003}_{-0.0003}$ & $-0.18^{+0.0002}_{-0.0003}$ & 3.23 & 3.12&4.67 \\ 
J16073773-3921388 & $10.20^{+0.209}_{-0.140}$ & $0.036^{+0.004}_{-0.004}$ & &&&$78.58^{+4.81}_{-5.1}$ & $101.20^{+5.05}_{-4.24}$ & $0.031^{+0.003}_{-0.003}$ & $-0.02^{+0.001}_{-0.001}$ & 0.62 & 6.04 &9.9\\ 
J16002612-4153553 & $10.88^{+0.083}_{-0.072}$ & $0.009^{+0.0009}_{-0.0008}$ && &&$53.11^{+8.81}_{-13.68}$ & $167.56^{+7.59}_{-11.63}$ & $-0.008^{+0.0006}_{-0.0006}$ & $-0.14^{+0.0007}_{-0.0007}$ & 0.63& 1.90 &2.85\\ 
Sz102 & $10.71^{+0.019}_{-0.018}$ & $0.056^{+0.0008}_{-0.0009}$ & &&&$78.50^{+0.57}_{-0.58}$ & $7.21^{+0.47}_{-0.48}$ & $0.20^{+0.0003}_{-0.0004}$ & $-0.098^{+0.0007}_{-0.0008}$ & 4.93 & 9.18&14.9 \\ 
Sz113 & $10.49^{+0.004}_{-0.004}$ & $0.049^{+0.0006}_{-0.0005}$ & &&&$26.02^{+1.63}_{-1.96}$ & $116.44^{+3.61}_{-4.21}$ & $0.032^{+0.0003}_{-0.0005}$ & $-0.23^{+0.0004}_{-0.0004}$ & 9.93 & 11.30&16.08 \\ 
Sz97 & $10.57^{+0.021}_{-0.021}$ & $0.024^{+0.001}_{-0.001}$ & &&&$55^{+2.71}_{-2.91}$ & $76.13^{+3.49}_{-3.75}$ & $0.059^{+0.0009}_{-0.0009}$ & $-0.18^{+0.0006}_{-0.0006}$ & 1.84 & 4.57&6.68 \\ 
J16085324-3914401 & $10.47^{+0.005}_{-0.005}$ & $0.051^{+0.0006}_{-0.0006}$ & &&&$48^{+0.96}_{-1.13}$ & $110.18^{+1.25}_{-1.15}$ & $0.006^{+0.0005}_{-0.0005}$ & $-0.15^{+0.0004}_{-0.0004}$ & 7.73 & 10.47&15.13\\ 
Sz77 & $10.63^{+0.025}_{-0.022}$ & $0.019^{+0.0009}_{-0.0009}$ & &&&$47.37^{+4.54}_{-5.31}$ & $110.79^{+5.52}_{-4.79}$ & $0.03^{+0.0008}_{-0.0007}$ & $-0.21^{+0.0007}_{-0.0005}$ & 1.65 & 3.78& 5.58\\ 
Sz130& $11.18^{+0.041}_{-0.034}$ & $0.008^{+0.0004}_{-0.0004}$ & &&&$37.73^{+7.4}_{-11.5}$ & $158.63^{+11.42}_{-17}$ & $0.056^{+0.0003}_{-0.0003}$ & $-0.16^{+0.0003}_{-0.0003}$ & 1.15 & 1.72&2.48 \\ 
Sz106 & $10.73^{+0.29}_{-0.16}$ & $0.024^{+0.002}_{-0.002}$ & &&&$81.31^{+4.37}_{-4.33}$ & $137.48^{+4.05}_{-4.3}$ & $-0.035^{+0.001}_{-0.001}$ & $-0.1^{+0.001}_{-0.001}$ & 0.69 & 3.87&6.54\\ 
V1192Sco & $10.65^{+0.16}_{-0.26}$ & $0.028^{+0.004}_{-0.004}$ & &&&$85.58^{+2.91}_{-4.43}$ & $157.64^{+4.81}_{-105.49}$ & $0.095^{+0.002}_{-0.002}$ & $-0.14^{+0.004}_{-0.004}$ & 0.39 & 4.21&6.79\\
\hline
\multicolumn{12}{c}{Binary Gaussian Models}\\
\hline
\hline
Sz81A& $10.78^{+0.01}_{-0.01}$ & $0.02^{+0.0004}_{-0.0006}$ & &&&$34.87^{+2.34}_{-2.34}$  & $133.55^{+5.45}_{-7.63}$   & $-0.26^{+0.0004}_{-0.0003}$  & $-0.87^{+0.0004}_{-0.0004}$  & 4.0 & 5.08&7.25\\ 
Sz81B&  $10.68^{+0.04}_{-0.04}$ & $0.02^{+0.001}_{-0.001}$ & &&&$59.08^{+4.55}_{-4.71}$& $128.86^{+7.73}_{-37.90}$ & $0.38^{+0.0008}_{-0.0007}$ & $0.95^{+0.0008}_{-0.0007}$ & 1.28 & 3.45&5.17 \\
Sz74A & $11.31^{+0.006}_{-0.006}$ & $0.02^{+0.0002}_{-0.0002}$ &&&&  $14.20^{+6.30}_{-8.12}$   & $36.03^{+17.10}_{-10.55}$   & $-0.07^{+0.0002}_{-0.0002}$  & $-0.43^{+0.0001}_{-0.0002}$ &7.94&3.90 &5.53\\
Sz74B& $10.76^{+0.01}_{-0.01}$ & $0.02^{+0.0006}_{-0.0006}$ &&&&$34.30^{+7.11}_{-10.00}$& $152.61^{+26.93}_{-86.56}$&$-0.09^{+0.0003}_{-0.0004}$&$-0.09^{+0.0004}_{-0.0004}$&3.20&4.67&6.65\\
V856ScoB & $11.21^{+0.007}_{-0.007}$ &$0.03^{+0.0003}_{-0.0003}$&&&&$64.34^{+0.56}_{-0.58}$&$133.66^{+0.66}_{-0.65}$ &$1.34^{+0.0002}_{-0.0002}$&$-0.56^{+0.0002}_{-0.0002}$&7.12&4.6&7.1 \\
\hline
\multicolumn{12}{c}{Binary Gaussian Model + delta function gaussian}\\
\hline
& & &  log I$_{\delta}$&  &  &&&  &   &  &\\
&&&[Jy sr$^{-1}$]&&&&&&&&\\
\hline
\hline
V856ScoA &  $11.40^{+0.004}_{-0.004}$ &$0.03^{+0.0002}_{-0.0002}$&$12.84^{+0.008}_{-0.009}$ &&&$57.24^{+0.27}_{-0.28}$ &$119.55^{+0.29}_{-0.29}$ &$0.005^{+0.00008}_{-0.00007}$ &$0.01^{+0.00006}_{-0.00006}$ &19.80&5.84&8.73 \\

\hline

\hline
\multicolumn{12}{c}{Gaussian with fix P.A}\\
\hline
\hline
J15450887-3417333& $10.49^{+0.005}_{-0.005}$ & $0.08^{+0.0007}_{-0.0007}$ &&&& $35.81^{+1.17}_{-1.23}$ & 0.0 & $0.008^{+0.0006}_{-0.0006}$ & $-0.08^{+0.0006}_{-0.0006}$ & 20.88 & 15.99&22.78 \\

\hline
\multicolumn{12}{c}{Gaussian with fix orientation}\\
\hline
\hline
\textsuperscript{a} J16075475-3915446 & $10.78^{+0.31}_{-0.21}$ & $0.006^{+0.002}_{-0.002}$ & &&&0.0 & 0.0 & $-0.0002^{+0.002}_{-0.002}$ & $0.006^{+0.001}_{-0.001}$ & 0.33 & 1.59&2.03 \\ 
\textsuperscript{a} J16084940-3905393 & $12.17^{+0.37}_{-0.45}$ & $0.002^{+0.001}_{-0.0007}$ & &&&0.0& 0.0 & $-0.03^{+0.0007}_{-0.0007}$ & $-0.23^{+0.0006}_{-0.0006}$ & 0.218 & 0.58&0.82 \\ 
\textsuperscript{a} J15592523-4235066 & $10.33^{+0.29}_{-0.20}$ & $0.010^{+0.003}_{-0.003}$ & &&&0.0 & 0.0& $-0.13^{+0.002}_{-0.003}$ & $-0.12^{+0.002}_{-0.003}$ & 0.30 & 2.26&3.14 \\

\hline

\multicolumn{12}{c}{Single Ring Models}\\

\hline

& &r$_{\rm ring}$ &  r$_{w}a$ & r$_{w}b$  &  &&&  &   &  &\\
&&["]&["]&["]&&&&&&&\\
\hline
\hline
Sz108B & $10.02^{+0.01}_{-0.01}$ & $0.08^{+0.005}_{-0.004}$ & $0.03^{+0.006}_{-0.005}$ & $0.06^{+0.002}_{-0.003}$ & &$54.87^{+0.60}_{-0.52}$ & $-21.04^{+0.66}_{-0.66}$ & $0.002^{+0.0008}_{-0.0008}$ & $0.02^{+0.0009}_{-0.0010}$ & 10.36& 18.41&25.12 \\ 
J16092697-3836269 & $10.19^{+0.34}_{-0.17}$ & $0.07^{+0.005}_{-0.006}$ & $0.01^{+0.008}_{-0.01}$ & $0.003^{+0.004}_{-0.002}$ & &$55.21^{+2.25}_{-2.65}$ & $-57.28^{+2.63}_{-2.35}$ & $0.13^{+0.001}_{-0.001}$ & $-0.10^{+0.001}_{-0.001}$ & 1.57 & 9.46&10.87 \\ 
Sz72 & $10.33^{+0.01}_{-0.01}$ & $0.05^{+0.003}_{-0.003}$ & $0.03^{+0.004}_{-0.004}$ & $0.01^{+0.002}_{-0.002}$ & &$31.46^{+1.60}_{-1.70}$ & $47.89^{+3.46}_{-3.44}$ & $0.007^{+0.0005}_{-0.0005}$ & $-0.18^{+0.0005}_{-0.0005}$ &5.41& 8.477&10.2\\ 
Sz90 & $10.17^{+0.01}_{-0.01}$ & $0.06^{+0.003}_{-0.002}$ & $0.002^{+0.003}_{-0.001}$ & $0.05^{+0.001}_{-0.002}$ & &$56.62^{+0.53}_{-0.55}$ & $135.87^{+0.75}_{-0.64}$ & $0.002^{+0.0006}_{-0.0007}$ & $-0.08^{+0.0006}_{-0.0006}$ & 7.36& 15.05&20.31\\ 
Sz96 & $10.59^{+0.30}_{-0.16}$ & $0.04^{+0.004}_{-0.004}$ & $0.005^{+0.005}_{-0.004}$ & $0.002^{+0.003}_{-0.001}$ & &$48.90^{+2.72}_{-3.43}$ & $23.34^{+3.65}_{-3.87}$ & $-0.004^{+0.0010}_{-0.0009}$ & $-0.15^{+0.0009}_{-0.0007}$ & 1.29 & 5.31&6.06 \\ 
Sz123A & $10.27^{+0.01}_{-0.01}$ & $0.20^{+0.0008}_{-0.0008}$ & $0.0002^{+0.0003}_{-0.0001}$ & $0.04^{+0.0008}_{-0.001}$ & &$53.34^{+0.17}_{-0.16}$ & $-25.20^{+0.28}_{-0.25}$ & $-0.006^{+0.0005}_{-0.0005}$ & $-0.12^{+0.0006}_{-0.0006}$ & 17.43 & 33.69&38.61\\ 
Sz100 & $10.03^{+0.008}_{-0.007}$ & $0.20^{+0.004}_{-0.002}$ & $0.03^{+0.003}_{-0.002}$ & $0.05^{+0.001}_{-0.002}$ & &$44.35^{+0.51}_{-0.23}$ & $67.28^{+0.34}_{-0.73}$ & $0.07^{+0.0007}_{-0.0008}$ & $-0.05^{+0.0006}_{-0.0007}$ & 24.47 & 33.39&39.61 \\ 
Sz131 & $10.15^{+0.03}_{-0.03}$ & $0.04^{+0.006}_{-0.005}$ & $0.01^{+0.008}_{-0.006}$ & $0.03^{+0.003}_{-0.003}$ & &$62.85^{+0.78}_{-0.84}$ & $-24.44^{+178.67}_{-2.36}$ & $0.003^{+0.0007}_{-0.0008}$ & $-0.21^{+0.001}_{-0.001}$ & 3.31& 9.74&13.78 \\

\hline
\multicolumn{12}{c}{Two gaussians}\\
\hline
& &r$_{c}$ & log I$_{0}b$ & r$_{\rm width}$ & r$_{\rm ring}$ &  &  &  &  &  &\\
&&["]&[Jy sr$^{-1}$]&["]&["]&&&&&&\\

\hline
\hline
Sz73 & $11.04^{+0.01}_{-0.01}$ & $0.03^{+0.0005}_{-0.0005}$ & $9.11^{+0.06}_{-0.03}$ & $0.06^{+0.005}_{-0.008}$ & $0.25^{+0.005}_{-0.008}$ & $42.76^{+1.63}_{-1.59}$ & $99.22^{+1.97}_{-4.28}$ & $-0.002^{+0.0003}_{-0.0003}$ & $-0.08^{+0.0003}_{-0.0003}$ & 12.56 & 28.32&41.57 \\
\hline
\label{tab:galariofit}
\end{tabular}%
}
\end{table*}

\begin{figure*}[!htb]
\centering
\includegraphics[width=2\columnwidth,trim={0cm 0.0cm 0cm 0.0cm}]{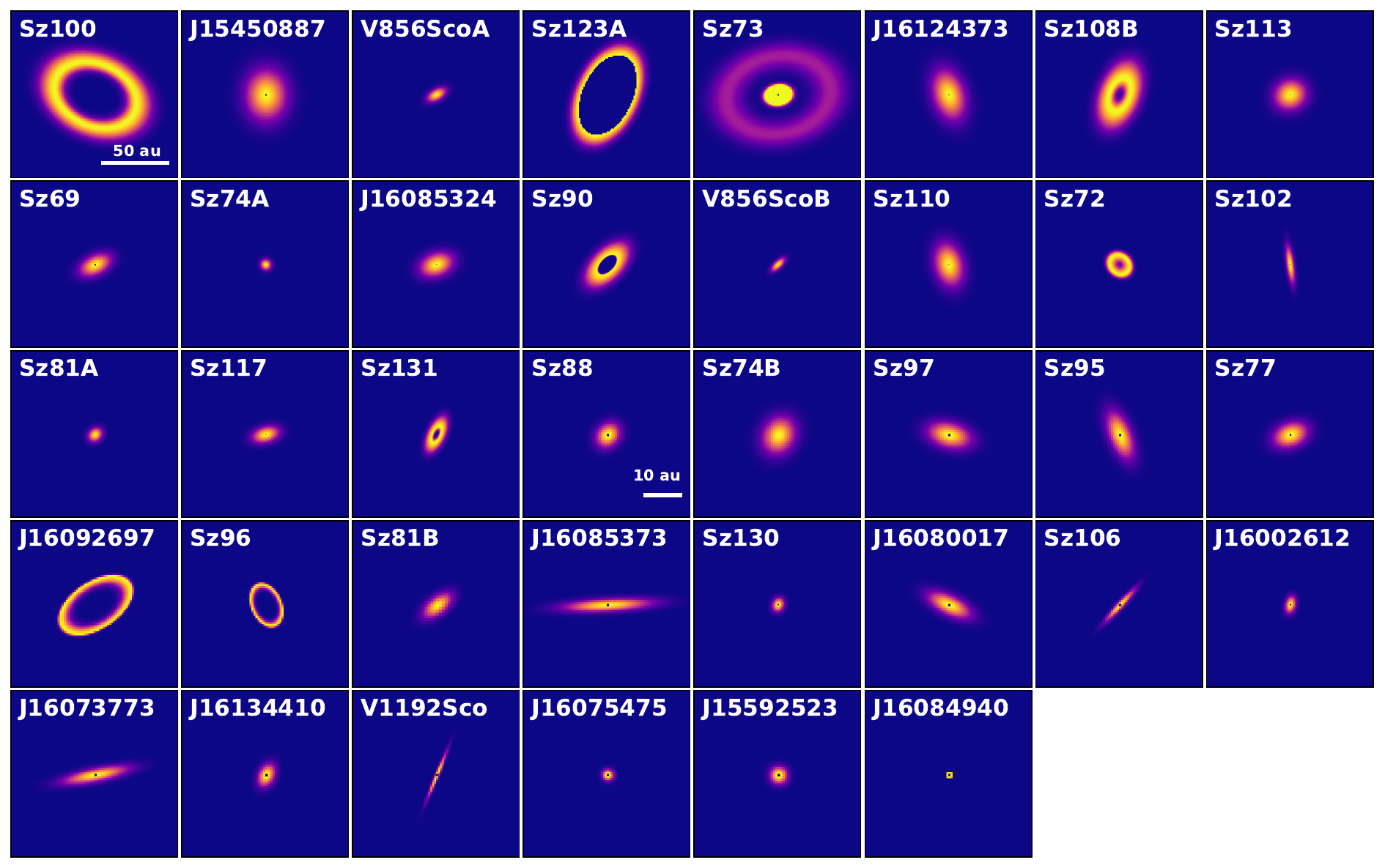}
\caption{\em{Best disk models obtained from the GALARIO visibility fitting. Note that the scale changes from Sz88 onwards to better highlight the most compact disks. The disks are arranged in descending order based on their total flux derived from the visibility fitting.}
}
\label{fig:galariomodels}
\end{figure*}

For most disks, the visibility fitting worked very well but there were a few cases which required specific adjustments.
The Gaussian model failed to capture the flux at the shortest baselines for Sz130, Sz77, and J16002612. In the case of Sz130, this resulted in a lower flux (41\% lower) compared to the literature values. For Sz77 and J16002612, the model appeared to overestimate the flux by 9,7\% and $\sim$3\%, respectively. It remains unclear whether these discrepancies are due to the low signal-to-noise ratio (SNR) of the observations or if a more complex emission pattern (potentially with a different orientation) is present that a simple Gaussian model cannot capture. For Sz130, the visibilities suggest the presence of some extended structure or emission that shows below 300 k$\lambda$, that we were unable to fit or determine its nature. Nevertheless, we opted to proceed with the model, as it represents only what we could directly observe. 

Three disks were unresolved or only marginally resolved, J16075475, J16084940, and J15592523. These disks exhibit extremely small sizes and very low surface brightness, leading to a low signal-to-noise ratio (SNR). Consequently, even with visibility fitting, the inclination and position angle (P.A.) could not be reliably determined. To improve the fitting of other parameters, the inclination and P.A. were fixed to face-on values. As a result, the radius measurements for these sources should be interpreted with caution. However, the overall fit remained satisfactory, and we retained these results for analysis and discussion in this paper. In the same case, we were unable to fit the position angle (P.A.) of J15450887-341733, which best fit was between 0.0 or 180 degrees. To obtain a plausible result, we fixed the P.A. to 0.0.

The faint ring in Sz73 was not detected during the initial imaging process with CASA, but became apparent when analyzing the visibilities.  However, we concluded that this is unlikely for many sources, as nothing was evident in the visibilities. Even the faint ring in Sz73, with surface brightness between 0.2 and 0.4 mJy, was detected above the 3-$\sigma$ level. Naturally, this depends on the observation brightness and the rms, but this means that fainter rings below this threshold would remain undetected.

In the Herbig star V856ScoA disk, a point source of strong emission is observed. 
The origin of this emission remains unknown and further discussion is needed. This emission, was detected at first only in the visibility plots, where the emission at long baselines never approaches zero, but it is observed in the continuum image. A model combining a Gaussian with a very narrow Gaussian successfully reproduced this feature, indicating that the emission is quite intense.  
Possible explanations include free-free emission from the central star's outflow or wind emission from the disk near the star. 

The Sz74 binary system shows a small substructure in the disk around its companion (Sz74B), potentially indicative of a cavity accompanied by a ring or asymmetry. However, due to insufficient resolution, it was impossible to constrain the size of the cavity or determine the disk's position angle, complicating the fitting process. To achieve a better fit, we opted to model the system using two Gaussians, as the substructure could also result from binary interactions, such as the formation of a vortex or a dust pile in the disk. While this approach improved the fit for the binary, the asymmetry remained unmodeled, and the position angle was still unconstrained, suggesting the source was very unresolved. Similarly, five additional disks (Sz113, Sz69, Sz103, Sz117) also show signs of cavities but were only fitted with a Gaussian profile since they were facing the same issue as Sz74B. Higher-resolution observations of these systems are necessary to confirm the origin of asymmetries, to better constrain the disk's orientations or the potential presence of a cavity.

\subsection{Disk size and substructure distribution}
\label{sec:disk_sizes}

\begin{figure}[!t]
\centering
\includegraphics[width=1.\columnwidth,trim={0cm 0.0cm 0cm 0.cm}]{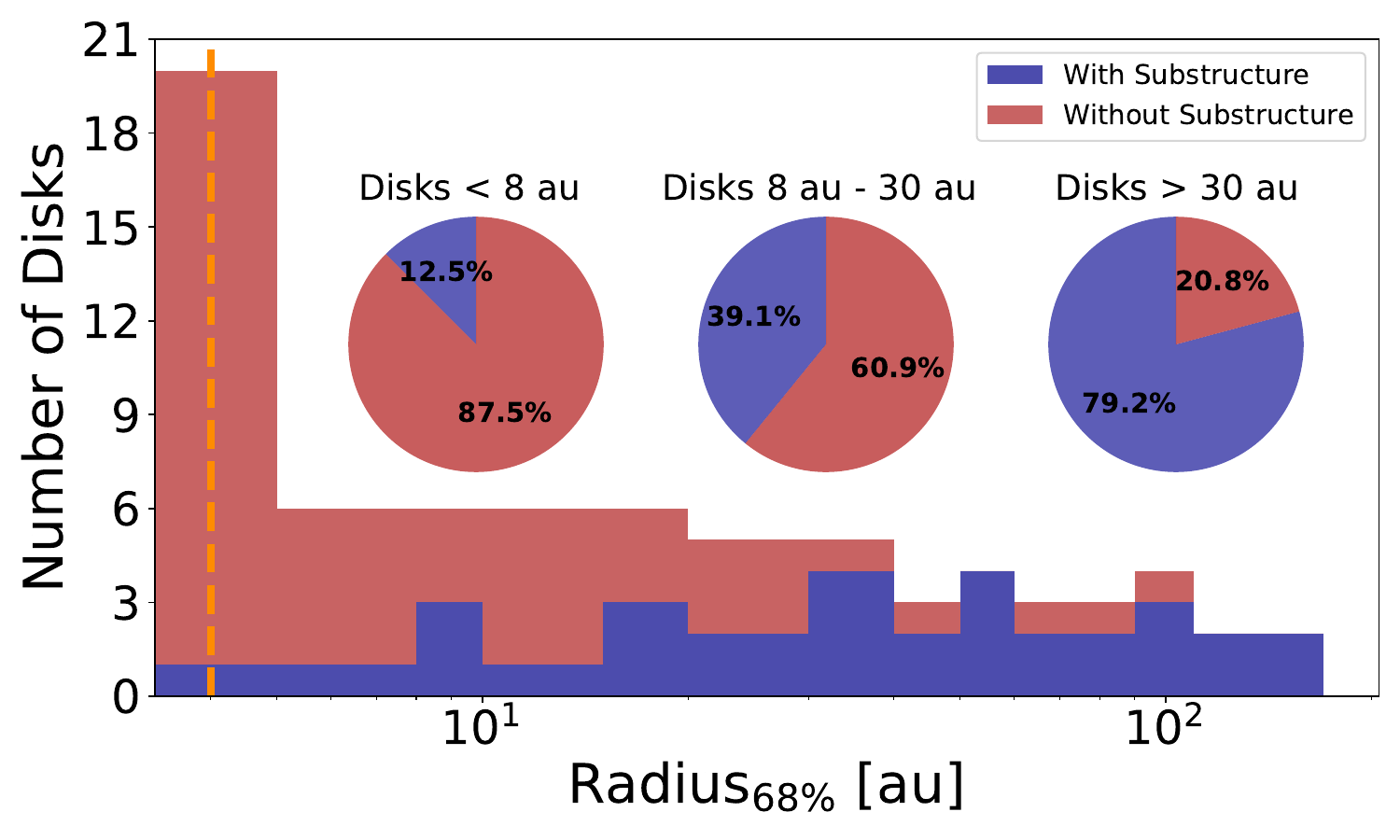}
\caption{\em{The size distribution of the Lupus disks. Disks exhibiting substructures in the continuum images are shown in dark blue, while smooth disks without visible substructures are depicted in red and added on top of the blue histogram. The pie charts inside the plot show the percentages of disks with and without substructures for sizes above and below 30 au. The vertical dashed orange line indicates our resolution limit, which is set at 0.03" (4 au). More than 67\% of the sample belongs to the compact disks classification.}
}
\label{fig:size_distribution}
\end{figure}

The \textsc{Galario} visibility fitting allows us to explore the size distribution of disks in the Lupus star-forming region at much smaller scales than previous works. We define disk sizes as the radius where the cumulative flux equals 68\% of the total flux (\citet{2018ApJ...865..157A}; \citet{2017ApJ...845...44T}).
The histogram in Figure~\ref{fig:size_distribution} shows a broad range of disk radii up to $\sim$160\,au and a marked increase at sizes smaller than 20\,au.

We find that two thirds of the disks are compact, with radii < 30 au. On the other hand, the majority of disks with large radii exhibit substructures, specifically at radii $\sim >30$ au, although they are relatively few in number. However, as we approach the resolution limit (indicated by the orange dashed line in Figure~\ref{fig:size_distribution}, the number of smooth disks increases significantly. This is expected, as we would not anticipate resolving any substructures at approximately three times the resolution limit, which is about 12 au. Consequently, while we can now better measure disk sizes and note an increase in the number of smaller disks, there remains insufficient resolution to discern whether a trend exists that correlates disk size with substructure occurrence. In fact, the more resolution we gain, the fewer smooth disks appear to show in the observations which points to a potential bias introduced by resolution where very small substructures may remain undetected in the smallest disks. For large disks with R$_{dust}>30$ au approximately 4/5 exhibit substructure detectable down to scales of 7 au. Conversely, in the compact disks, at most 72\% (5 out of 7) are smooth, with substructures down to scales of 4 au. Considering all disks in the sample, the data suggests that about 3 out of 8 exhibit substructure, regardless of disk size.

\begin{figure}[!t]
\centering
\includegraphics[width=1\columnwidth,trim={0cm 0.0cm 0cm 0.0cm}]{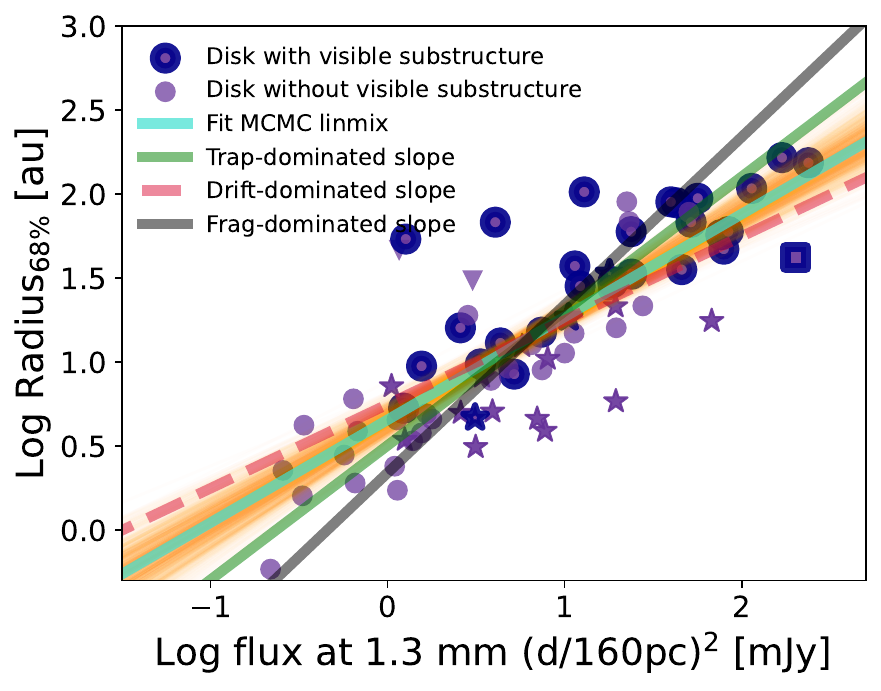}
\caption{\em{The size-luminosity relation (SLR) for the Lupus star-forming region at high resolution. Disks with substructures are marked by purple circles outlined in dark blue, while smooth disks are represented by solid purple circles without outlines. Unresolved disks are indicated by downward-pointing triangles. Binary disks are distinguished by star-shaped markers, and RU Lup is identified with a square marker for clarity. The drift-dominated slope, as described by \citet{2019MNRAS.486L..63R}, is illustrated by the dashed red line, the trap-dominated slope \citep{2022A&A...661A..66Z} is shown in green and the frag-dominated slope in black. Our fit, using the \textsc{linmix} package, is displayed in turquoise. All disks are normalized to a common distance of 160 pc for consistency.
}}
\label{fig:SLR}
\end{figure}

We plot the disk radii versus their total flux density (scaled to 160\,pc) in Figure~\ref{fig:SLR}. There is a clear relationship between the two observables and we apply a Bayesian linear regression method implemented in the \textsc{linmix} package (Joshua E. Meyers, \citet{2007ApJ...665.1489K}),
\begin{equation}
\log_{10}(R_{68}) =\alpha+\beta\log_{10}(F_{\rm mm}).
\end{equation}
The fitted values are an intercept (normalization) $\alpha = 0.66 \pm 0.06$ and slope (power law index) $\beta = 0.61 \pm 0.06$, with an intrinsic dispersion of $0.116 \pm 0.02$

This relation can constrain if dust evolution is dominated by drift or by traps being present in the disk (\citet{2019MNRAS.486L..63R}; \citet{2022A&A...661A..66Z}). The observed slope falls between the drift-dominated and the trap- and frag-dominated regimes which is consistent with previous findings by \citet{2018ApJ...865..157A} and \citet{2020ApJ...895..126H}.

Notably, a few disks lie significantly below the SLR of Lupus. Most of these disks are in binary systems (star markers in Fig. ~\ref{fig:SLR}), and the interaction between companions likely truncates their radius, as discussed in \citet{2022A&A...662A.121R}, affecting their position on the SLR. Another exception is RU Lup (square in Fig. ~\ref{fig:SLR}), one of the most active T Tauri stars, which likely experiences increased flux caused by its strong accretion features (e.g \citet{2008A&A...482L..35G}; \citet{2016MNRAS.456.3972S}), placing it below the SLR as well.
Among the five disks that appear high above in the SLR, two (J16102955-3922144 and J16000060-4221567) have only been observed at low resolution \citep{2018ApJ...859...21A}, therefore, their radius might be overestimated. The other three disks have large cavities and may be more evolved. As mentioned by \citet{2022A&A...661A..66Z}, the presence and location of planets within a disk can alter the SLR by influencing the disk’s evolutionary path. Given that the SLR represents only a snapshot of a cluster, with a diverse mix of stellar masses and angular momenta, it remains challenging to draw definitive conclusions about the outliers.

\subsection{Radiative transfer modeling}
Whereas disk dust masses are usually calculated using a linear relation with the millimeter-flux \citep[e.g.][]{2018ApJ...859...21A}, this relation relies on the assumption that the dust continuum emission is optically thin. The existence of very small disks raises questions about how well we can determine their disk masses as the optical depth may be very high. We therefore created a large grid of models to calculate the expected total flux densities for different stellar parameters and disk masses spanning the range of observed radii.
We performed radiative transfer at 1.3~mm for the generic protoplanetary disk model in the RADMC-3D software package \citep{2012ascl.soft02015D} with a dust density distribution,
\begin{eqnarray}
\rho(r,z) =\frac{\Sigma(r)}{\sqrt{2\pi}H_p}\exp\left(-\frac{z^2}{2H_p^2}\right),
\label{eq: Density distribution}
\end{eqnarray}
where $r$ represents the radial distance from the star, $H_p(r)$ is the disk vertical scale height, and $\Sigma(r)$ is the dust surface density,
\begin{eqnarray}
\Sigma(r)= \Sigma_0\left(\frac{r}{r_{\rm out}}\right)^{-1},
\label{eq: Surface Density distribution}
\end{eqnarray}
defined out to an outer radius $r_{\rm out}$.

We produced a grid of 1728 models, varying three key parameters: 12 disk dust masses, 12 radius, 12 stellar luminosities (L$_{\odot}$). 
These parameters ranged from 4e-6 to 5.1e-3 $M_{\odot}$, 0.5 to 30 au, 0.0025 to 3 $L_{\odot}$, with values spaced in a log-uniform manner, respectively. 
Unsurprisingly, the star’s effective temperature has minimal impact on the flux-derived dust mass of the models as the absorbed stellar radiation is fully reprocessed by the disk to an equilibrium that is determined only by the energy input and dust distribution, therefore, we did not include it in the parameter space. 
The parameter limits were based on \citet{2017A&A...600A..20A} for Lupus sources, later updated using Gaia DR2 distance corrections as described in \citet{2019A&A...629A.108A}. The stellar parameters of the Lupus sources, where known, are provided in Table~\ref{table:Stellar parameters} and Table~\ref{table:Stellar parameters 2}.

To calculate the dust opacities needed inside RADCM3D, we utilized the \texttt{optool} software \citep{2021ascl.soft04010D} with DSHARP opacities \citep{2018ApJ...869L..45B}. The dust opacity values were calculated using grain sizes from $a_{\text{min}} = 0.050  \mu$m to $a_{\text{max}} = 3$ mm, incorporating the full scattering matrix.

For each model, we create a ray-traced image assuming a distance of 160\,pc and face-on disk inclination and measure the total flux and $R_{68}$ radius in the same way as for the observations.

\begin{figure}[!t]
\centering
\includegraphics[width=1.02\columnwidth,trim={0cm 0.0cm 0cm 0.0cm}]{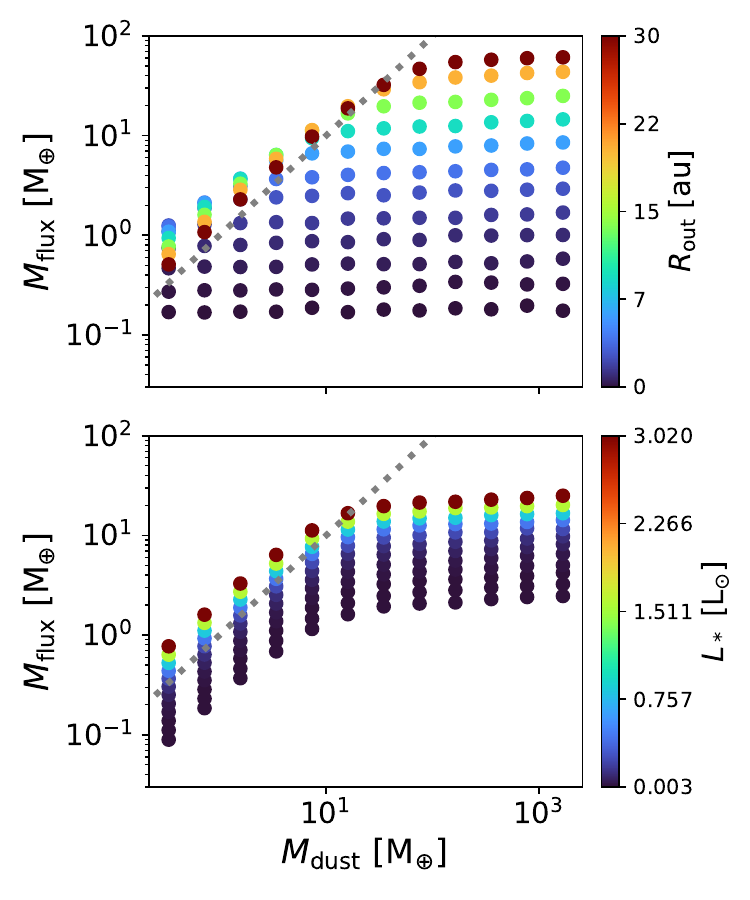}
\caption{\em{Comparison of the disk dust mass from a grid of 1728 models versus the mass derived from flux for the same models. In the top panel, stellar luminosity and effective temperature are fixed, illustrating how disk radius varies with flux and dust mass, with the radius represented by different colors on the colorbar. The bottom panel shows the effect of changing stellar luminosity while keeping disk radius and effective temperature constant. The gray dotted line highlights the masses from the flux that equal the disk dust mass introduced in the models.
}}
\label{fig:mass_comparison}
\end{figure}

We first compare the disk masses as derived from the simulated flux, $M_{\rm flux}$, with the model input, $M_{\rm model}$, in Figure~\ref{fig:mass_comparison}. For the former, we applied the optically thin approximation for dust mass at 1.3 mm simplified in \citet{2018ApJ...859...21A},
\begin{equation}
M_{\text{flux}} = \frac{F_{\nu} d^2}{\kappa_{\nu} B_{\nu}(T_{\text{dust}})}
                  \simeq 0.68\left(\frac{d}{160 \text{pc}}\right)^2\left(\frac{F_{1.33\text{mm}}}{\rm mJy}\right)\,M_\oplus
\end{equation}
where the dust opacity, $\kappa_{\nu}=2.3\,{\rm cm}^{2}\,{\rm g}^{-1}$ \citep{1990AJ.....99..924B}, $B_{\nu}$ is the Planck function, and we assume a uniform dust temperature, $T_{\rm dust} = 20$ K.

Figure~\ref{fig:mass_comparison} primarily highlights the saturation of the disk flux as the emission becomes optically thick, demonstrated by how the models deviate from the dashed gray line that indicates equality of $M_{\rm flux}$ and $M_{\rm dust}$.
Indeed, the flux-derived mass scales in a systematic way with the stellar luminosity for each input mass. Small disks saturate at lower disk masses than larger disks due to their much higher average optical depths.

Motivated by this behavior, we fit power laws to the model grid away from the saturated region, defined through the mean optical depth,
\begin{equation}
\bar\tau = \kappa\bar\Sigma = \frac{\kappa M_{\rm dust}}{\pi R_{\rm out}^2}
         \simeq 19.4\left(\frac{M_{\rm dust}}{M_\oplus}\right)\left(\frac{R_{\rm out}}{{\rm au}}\right)^{-2}
\label{eq:mean_tau}
\end{equation}
which holds independently of the surface density profile and the normalization is for our mean observing wavelength, 1.33\,mm.
For very optically thin emission, $\bar\tau < 0.25$,
\begin{equation}
M_{\rm flux} \simeq 7 \left(\frac{R_{\rm out}}{1\,{\rm au}}\right)^{-0.5} \left(\frac{L_\ast}{1\,L_\odot}\right)^{0.3}\,M_{\rm dust},
\label{eq:lowtau_mass}
\end{equation}
to within 10\%.
As $\bar\tau$ increases, the behavior begins to deviate from a linear dependence on $M_{\rm dust}$ and is a more complicated function of radius and luminosity.
For optically thick emission, $\bar\tau > 1$, the mass dependence is so weak that the flux is no longer a reliable measure of dust mass. This presents a chicken and egg problem because we do not know how accurately we can determine the disk mass without knowing the optical depth and vice versa. However, the size-luminosity relation provides a way forward.

\subsection{Dust mass estimates}
\begin{figure}[!t]
\centering
\includegraphics[width=1\columnwidth]{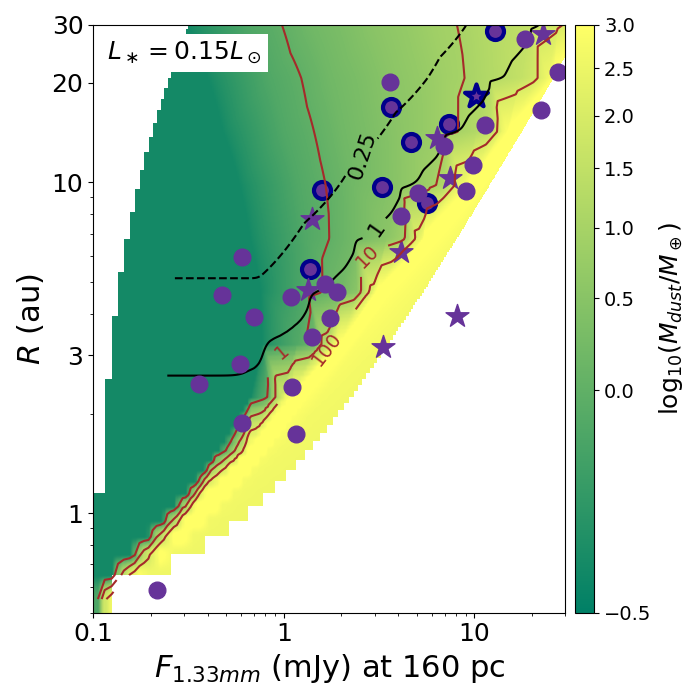}
\caption{\em{Comparison of the model grid with observed disk radii and total flux (normalized to 160\,pc). The image represents the model dust mass on a log scale (also in brown contours) and the black contours show the mean optical depth, $\bar\tau$, defined in equation~\ref{eq:mean_tau}, with values of 0.25 and 1. The dark blue edges indicate a visible substructure in this source and a star-shaped marker distinguishes binary disks.
\label{fig:flux_radius}
}}
\end{figure}

As shown in \S\ref{sec:disk_sizes}, our survey of compact disks extends the size-luminosity relation to much lower disk radii than previous work. For a given stellar luminosity, our model grid calculates the flux for a given disk mass and radius. We then invert this relationship to map the disk mass on the flux-radius plane in Figure~\ref{fig:flux_radius}, which allows a direct comparison with the two principal observables in our survey.

The locus of the model shifts horizontally with the stellar luminosity but in principle, this allows the disk mass to be estimated from our model. Relatively large, faint disks have low $\bar\tau$ and the millimeter flux scales with the mass. However, disks that are relatively bright for their size may be optically thick and very massive. A handful of outliers that lie on or beyond the yellow region of this plot are binaries with relatively low radii for their flux due to tidal truncation and are not well represented by our models.
However, most disks lie in the regime of intermediate optical depth, $0.25 < \bar\tau < 1$, where radiative transfer modeling is necessary to determine their dust mass.

\begin{figure}[!t]     
\centering
\includegraphics[width=1.\columnwidth,trim={0cm 6.cm 0cm 0.0cm}]{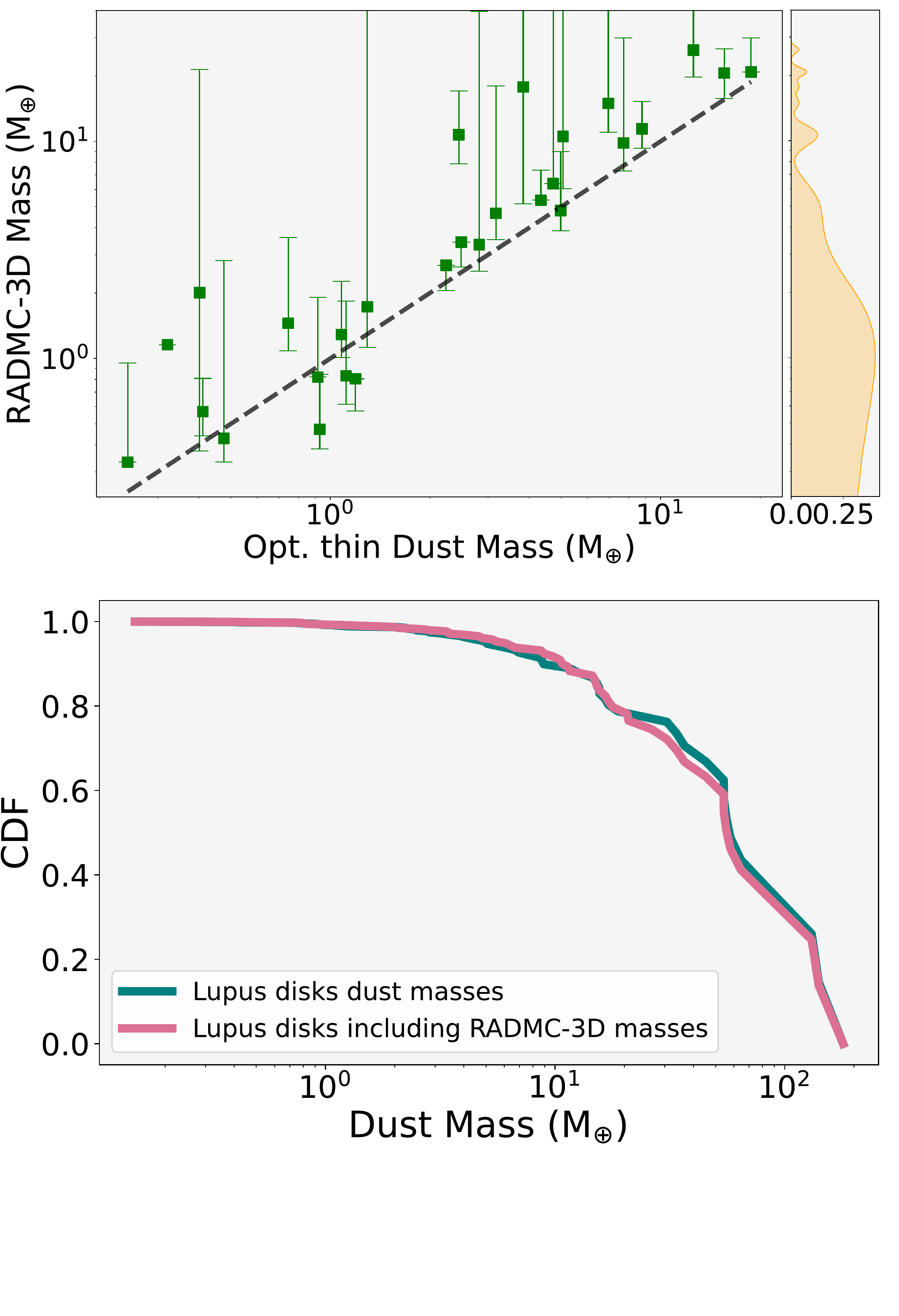}
\caption{\em{Top panel: Masses of the compact disks in Lupus (R < 30 au). These masses were obtained through the RADMC-3D model grid for $\tau$ < 5. The black dashed line indicates where the RADMC-3D dust mass equals the optically thin dust mass calculation, eq. 10.  Top right panel: Kernel density estimation of the disk dust masses. Bottom panel: Cumulative distribution functions (CDFs) of all the disk masses in Lupus, using two approaches: in teal, all masses are estimated with the optically thin approximation based on flux values, eq. 10, and in pink, we substitute the masses with values derived from our model grid where available.}
}
\label{fig:diskdustmasses}
\end{figure}

The observations extend in the same direction as the $\bar\tau$ contours. For small optical depths, the flux is proportional to mass so, based on equation~\ref{eq:mean_tau}, this is similar to the relation found by \citet{2018ApJ...865..157A} who found that the millimeter luminosity scales with the disk surface area. They suggested a possible explanation in which the emission comes from optically thick substructures with a filling fraction of $\sim 0.3$. 

\begin{figure}[!t]
\centering
\includegraphics[width=1.02\columnwidth]{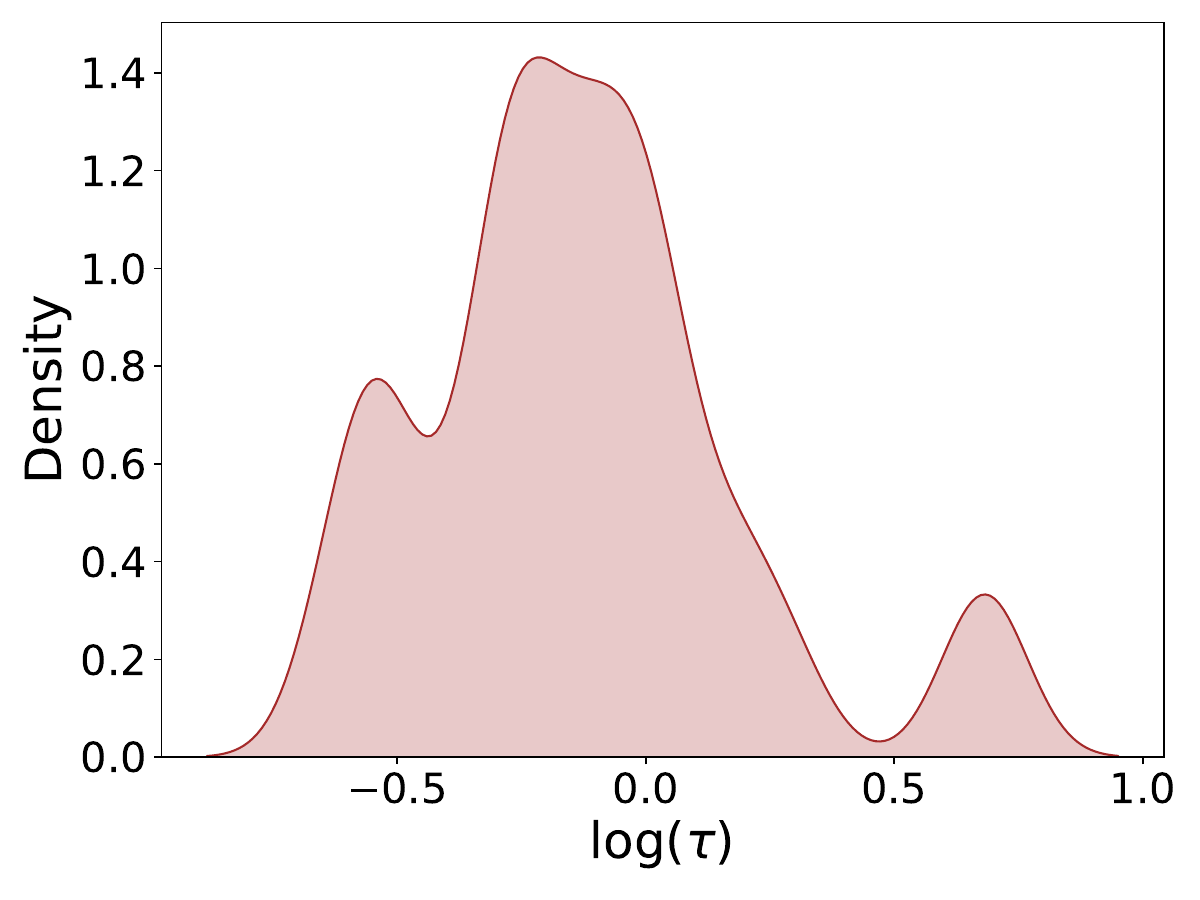}
\caption{\em{Optical depths of the Lupus disks using a kernel density estimation for points with radii < 30 au and obtained through Eq.~\ref{eq:mean_tau} and the grid measurements. }
}
\label{fig:tausdisks}
\end{figure}

Using the stellar luminosities tabulated in Appendix~\ref{appendix:a}, we can create a bespoke radius-flux grid for each disk and thereby determine its dust mass. The uncertainties in each observable are readily propagated through. Several disks lay at the edge of their grids (Sz74A, Sz113, Sz102, J16084940-3905393, J15450887-3417333, Sz81A,  Sz112, Sz88, Sz69, Sz130, HT LupA, Sz104, J16002612-4153553, J16085373-3914367), due to either being part of a binary system, having poorly constrained luminosity, or being highly optically thick with $\bar\tau \gg 5$, resulting in inferred masses that were either indeterminate or exceptionally large with substantial uncertainties. The distribution of dust masses for the 28 disks with well determined values and known luminosities and the cumulative distribution of all Lupus disks are presented in Figure~\ref{fig:diskdustmasses} and Table~\ref{tab:Disk dust masses}. Additionally, the $\tau$ values are plotted in Figure~\ref{fig:tausdisks}.

\subsection{The exoplanet population in the Lupus substructures}

To gain insight into the potential exoplanet population within the Lupus protoplanetary disk sample, we estimated the planetary masses that could dynamically cause each of the observed gaps.
We employed DBNets, a deep-learning tool that utilizes convolutional neural networks (CNNs) to analyze observations of dust continuum emission and predict the mass of the gap-opening planets, as developed by \citet{2024A&A...685A..65R}. DBNets takes several inputs: the continuum emission image of the disk (in the form of a FITS file), the center of the image (in pixels), the disk's orientation (inclination and position angle), the distance to the source, and the position of the planet (in astronomical units), which corresponds to the center of the gap being studied. In total, we analyzed 25 substructures within our sample. For well-studied disks such as RULup, IMLup, Sz129, Sz114, and GWLup, we used the planet masses already estimated in \citet{2024A&A...685A..65R}.

For the remaining 16 disks and 17 substructures, we ran the code using the continuum FITS files provided in Section 2 (11 from the new images and 5 from other projects). For cavity positions we used $\frac{r_{\rm ring}-r_{w}a}{2}$ while for the only gap, we used $\frac{r_{\rm ring} - r_{\rm width}+r_{c}}{2}$. Inclinations and position angles, which were not obtained in Section 3 of this paper, were sourced from previous studies. Specifically, for Sz98, we adopted values from \citet{2023A&A...679A.117G}, for J16083070-3828268 from \citet{2019A&A...624A...7V}, for Sz91 from \citet{2021ApJ...923..128M}, and for Sz76 and J16090141-3925119 from \citet{2022arXiv220408225V}. For RYLup and Sz111, we referred to \citet{2018ApJ...854..177V}. 

\begin{table}[ht]
\centering
\caption{Dust masses and optical depths for a subsample of 28 Lupus disks with radii < 30 au, which fit within our model grid.}
     \setlength\tabcolsep{3.pt}
     \renewcommand{\arraystretch}{1.35}
\begin{tabular}{ccc}
\hline

Source & Dust Mass [M$_{\oplus}$] & $\tau$ \\
\hline
\hline
J16124373-3815031   & 9.8$^{+20.1}_{-2.5}$ & 0.86   \\
Sz117        & 3.3$^{+36.3}_{-0.8}$ & 1.03   \\
Sz110        & 6.4$^{+92.6}_{-0.0}$ & 0.75    \\
J16134410-3736462   & 2.0$^{+19.4}_{-1.6}$ & 5.00  \\
J16080017-3902595   & 1.5$^{+2.1}_{-0.4}$ & 1.40  \\
Sz95         & 0.8$^{+1.0}_{-0.2}$ & 0.67  \\
J16073773-3921388   & 0.6$^{+0.2}_{-0.1}$ & 0.31   \\
J16085324-3914401   & 10.5$^{+1030.6}_{-4.5}$ & 1.93\\
Sz97&1.7$^{+176.7}_{-0.6}$ & 1.56\\
Sz77         & 0.8$^{+0.0}_{-0.2}$ & 1.03  \\
Sz106        & 0.4$^{+2.4}_{-0.1}$ & 0.54 \\
V1192Sco     & 1.2$^{+0.0}_{-0.0}$ & 1.07   \\
Sz81B        & 0.8$^{+1.1}_{-0.0}$ & 0.72    \\
J15592523-4235066   & 0.3$^{+0.6}_{-0.0}$ & 1.07  \\
Sz108B       & 14.9$^{+961.3}_{-3.9}$ & 0.87  \\
J16092697-3836269   & 1.3$^{+1.0}_{-0.3}$ & 0.28 \\
Sz72         & 17.8$^{+47.3}_{-12.6}$ & 4.63  \\
Sz90         & 4.8$^{+4.2}_{-0.9}$ & 0.41  \\
Sz96         & 0.5$^{+0.4}_{-0.1}$ & 0.31   \\
Sz131        & 2.7$^{+0.0}_{-0.6}$ & 0.55   \\
Sz66         & 5.4$^{+2.0}_{-0.0}$ & 0.56   \\
Sz65&20.6$^{+6.1}_{-4.9}$ & 0.51\\
Sz76         & 3.4$^{+0.0}_{-0.8}$ & 0.23  \\
Sz103        & 4.7$^{+13.3}_{-1.1}$ & 0.52  \\
J16081497-3857145   & 10.7$^{+6.3}_{-2.8}$ & 0.51  \\
Sz73&11.4$^{+3.8}_{-2.1}$ & 0.27\\
GQLup&20.9$^{+9.1}_{-0.0}$ & 0.87\\
RXJ1556.1-3655&26.3$^{+15.7}_{-6.5}$ & 0.69\\

\hline
\end{tabular}
\label{tab:Disk dust masses}
\end{table}

\begin{figure*}[!htb]
\centering
\includegraphics[width=2.05\columnwidth,trim={0.0cm 0.0cm 0cm 0.0cm}]{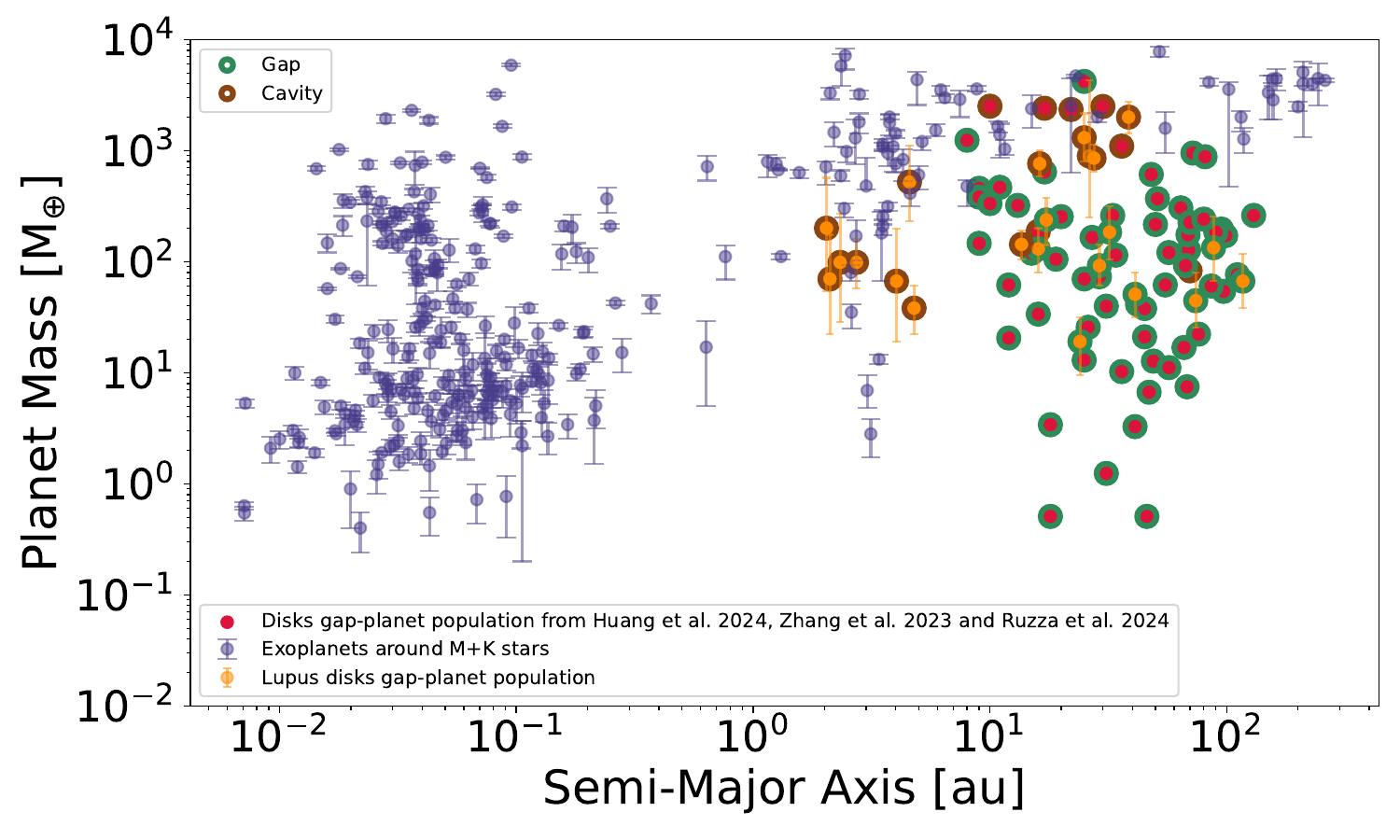}
\caption{\em{Lupus exoplanet masses obtained using DBNets plotted against semi-major axis, assuming that the gaps are carved by planets. In blue, we show all observed exoplanets around M and K stars from the NASA Exoplanet Archive catalog, while the orange markers represent estimates for the Lupus sample. Brown outlines indicate cavities as the type of substructures, while green outlines denote gaps and rings. Additionally, planet mass estimates from \citet{2023ApJ...952..108Z}, \citet{2024arXiv241003823H}, and \citet{2024A&A...685A..65R} are highlighted in crimson circle markers.}}
\label{fig. Dbnets-planets}
\end{figure*}

\begin{figure}[!htb]
\centering
\includegraphics[width=1.0\columnwidth,trim={0.0cm 0.0cm 0cm 0.0cm}]{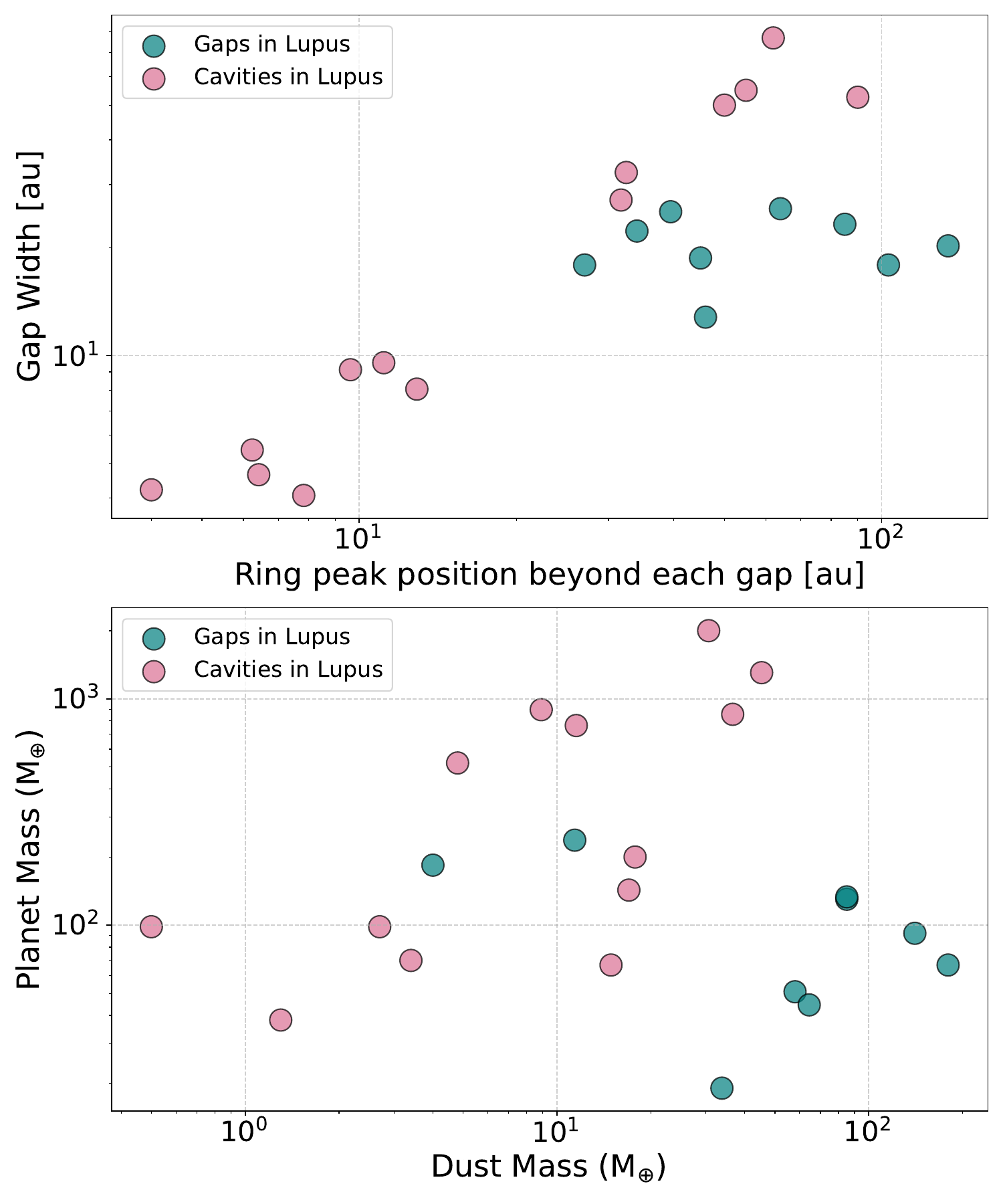}
\caption{\em{Top panel: Gap widths and locations of the ring peaks beyond each gap in our disks, as well as the DSHARP disks \citep{2018ApJ...869L..47Z} in Lupus. Gaps are shown in teal, while cavities are highlighted in pink. Bottom panel: Disk dust masses versus planet masses from the DBNets analysis, using the same colors to denote the substructure type. }}
\label{fig.massvsmassgaps}
\end{figure}

The results, obtained for a range of $\alpha$-turbulence values between $10^{-2} - 10^{-4}$, including predicted planet masses,  their radial positions, and whether the uncertainties exceed DBNet's rejection threshold, are summarized in Table ~\ref{tab:planet_masses}. It is important to note that this rejection threshold defines the significance of the results, meaning these values should be interpreted with caution.  Additionally, in Figure  ~\ref{fig. Dbnets-planets}, we show the full population of known exoplanets around M and K stars (sourced from the NASA Exoplanet Archive in November 2024) alongside our predicted exoplanet population in the Lupus star-forming region. Most exoplanets in the archive were discovered via Radial Velocity (RV), providing minimum mass values (Msini), and Transit Timing Variations (TTV), which yield more precise masses when combined with RV data. The disk dust masses obtained in the previous section, where applicable, and the dust masses derived from the flux for the larger disks are plotted at the bottom of Figure~\ref{fig.massvsmassgaps} together with the planet masses for comparison. Moreover, we show the ring peak position beyond each gap and gap widths (this work and the DSHARP disks, \citet{2018ApJ...869L..47Z}) at the top of Figure~\ref{fig.massvsmassgaps}.

\begin{table}[h!]
    \centering
    \caption{Gap centers and planet mass estimates for the subsample of Lupus disks with substructures. Of these, 11 are from our new observations, while 5 were obtained from FITS files of other projects. The final 5 disks are the values taken from \citet{2024A&A...685A..65R}. \textsuperscript{*} Due to the high uncertainties exceeding DBNet's rejection threshold, the planet masses for these disks should be interpreted with caution.}
     \setlength\tabcolsep{5pt}
     \renewcommand{\arraystretch}{1.3}
    \begin{tabular}{ccc}
        \hline
        Source & Gap center [au] & Planet Mass [M$_{\oplus}$]\\
        \hline
        \hline
        \textsuperscript{*}Sz108B                 & 4.03   & 66.7$^{+130.6}_{-47.6}$ \\
        J16092697-3836269       & 4.78   & 38.3$^{+22.3}_{-15.8}$ \\
        \textsuperscript{*}Sz72                   & 2.04   & 200.0$^{+370.5}_{-146.8}$ \\
        \textsuperscript{*}Sz90                   & 4.57   & 520.6$^{+581.5}_{-289.5}$ \\
        Sz96                   & 2.73   & 98.5$^{+60.6}_{-41.5}$ \\
        Sz123A                 & 16.22  & 763.8$^{+251.2}_{-181.6}$ \\
        Sz100                  & 13.59  & 143.5$^{+47.8}_{-38.4}$ \\
        \textsuperscript{*}Sz131                  & 2.33   & 98.5$^{+171.2}_{-70.2}$ \\
        Sz73                   & 17.3 & 237.6$^{+139.9}_{-96.3}$ \\
        J16083070-3828268       & 38.5  & 2008.3$^{+743.8}_{-566.5}$ \\
        Sz98                   & 16  & 130.3$^{+70.1}_{-51.0}$ \\
        \textsuperscript{*}Sz98                   & 88  & 133.4$^{+121.7}_{-67.5}$ \\
        \textsuperscript{*}Sz76                   & 2.11   & 70.2$^{+105.7}_{-46.8}$ \\
        RYLup                  & 25  & 1306.0$^{+854.7}_{-511.6}$ \\
        \textsuperscript{*}Sz91                   & 26.3  & 897.6$^{+3413.6}_{-646.5}$ \\
        J16090141-3925119       & 32  & 184.6$^{+133.5}_{-79.6}$ \\
        Sz111                  & 27.5  & 855.4$^{+274.0}_{-213.8}$ \\
        Sz129                  & 41  & 50.4$^{+28.7}_{-19.0}$ \\
        RULup                 & 29  & 92.3$^{+50.8}_{-31.8}$ \\
        IMLup                  & 117 & 66.9$^{+51.4}_{-27.0}$ \\
        Sz114                  & 24  & 19.1$^{+12.4}_{-9.1}$ \\
        GWLup                  & 74  & 44.6$^{+35.1}_{-20.0}$ \\
        \hline
    \end{tabular}

    \label{tab:planet_masses}
\end{table}
From Table ~\ref{tab:planet_masses} we see that the range of inferred planet masses in Lupus, $\sim 20-2000\,M_\oplus$, is comparable to similarly derived estimates from observations of disks in Taurus \citep{2023ApJ...952..108Z} and the $\sigma$ Orionis cluster \citep{2024arXiv241003823H}. In addition, our high resolution images reveal gaps at such small radii that the inferred planets overlap with the cluster of radial velocity detected exoplanets.

\section{Discussion}
\label{sec:discussion}

\subsection{New insights into the substructures and origin of compact disks}
Our high-resolution compilation of the Lupus disk population shows a large number of compact disks with a relatively small fraction of substructures, whereas substructures are common in large disks with radii $> 30$ au, with 79\% of them displaying rings or gaps (19 out of 24). However, since the majority of Lupus disks are compact, this fraction may not be representative of the entire disk population. 
Many previous studies have shown that initially featureless disks reveal greater complexity when the resolution is increased (\citealt{2019ApJ...882...49L}; \citealt{2021A&A...645A.139K}; \citealt{2022arXiv220408225V}; \citealt{2024ApJ...966...59S}).
Here, we have detected new substructures, mainly cavities down to 4 au in radius, but only in 24\% of the compact disks, excluding those that remain unresolved even in the visibility analysis.

Most of the structured compact disks show cleared cavities rather than rings, with the notable exception of Sz73. At lower resolution, \citet{2018ApJ...854..177V} found that 15\%\footnote{This is recalculated (11/73) for the total number of confirmed Lupus members: \citet{2018ApJ...854..177V} reported 11\% (11/96).}  
of the Lupus disk population show large cavities of $>20$ au radius. Our high resolution survey increases the resolved transition disk fraction to 24.6\% (18 out of 73), including cavities down to 4 au radius. 

Sz73 is the sole structured compact disk with a ring. This is faint, between 0.2 and 0.4 mJy, but should be detectable in most of the other disks in the survey. Furthermore, by utilizing \textsc{Galario}, we reduced analysis biases and are therefore reasonably confident that there are few, if any, other Sz73-like objects in Lupus.

Our results thus confirm that many of the unresolved disks from \citet{2018ApJ...859...21A} are in fact very compact, consistent with the hypothesis from \citet{2021AJ....162...28V}. They speculate that disks which have not formed giant planets early on to halt the pebble drift, will rapidly become compact, following dust evolution models \citep[][Appelgren et al. subm.]{Pinilla2020}. Subsequently, \citet{2024A&A...689A.236S} demonstrated that such drift-dominated disks are capable of producing the close-in Super-Earth population around M-dwarfs through pebble accretion.

The observed properties of the compact disks in Lupus are consistent with this scenario. Most of our compact disks (73\%) are found around M-type stars, with only 14\% around K-type stars (The rest don't have a determined Spectral type). However, this distribution reverses in larger disks: 50\% are around M-type stars, but already 41\%  hosted by K-type stars. Furthermore, \citet{2024A&A...689A.236S} showed that the disk dust masses in their models are $\sim$ few $M_{\oplus}$ and a few au in size after 1-2 Myr due to the inward drift in combination with inner planet formation through pebble accretion. The observed dust masses are in that case the remnant of planet formation and not representative for the solid mass budget for planet formation. A fraction of the formed planets in their simulations reach pebble isolation mass, creating small dust traps, which may be the explanation for some of the observed small cavities in compact disks. While forming wide gaps in the first 10 au of the disk is challenging, \citet{2024A&A...692A..45K} proposed that low-mass planets could carve such gaps in the inner disk (<10 au), with widths as large as their position. For compact disks without observed substructure, either planets have not reached pebble isolation mass or they are located so close-in that the dust traps remained unresolved. Predictions of the exact scenarios for individual compact disks will be explored in a follow-up study (Guerra-Alvarado et al. in prep.). 

Also in the sample of large, well-resolved disks in Lupus there remain several disks that do not show substructure ($\sim$25\%  of the disks with radius > 15 au) (e.g J16011549, J16000236, Sz133, Sz65, RXJ1556.1, GQLup, J16081497, J15450887, HTLupA). \citet{2023A&A...673A..77R} identified a disk around MP Mus without any substructure at 4 au resolution and proposed that in such evolved systems, large grains should have drifted onto the star unless some mechanism was preventing this. This means that substructures could remain undetected due to high optical depth at 1.3 mm or the substructures could be smaller than the current resolution limits of the observations. This could also be the case for some of the disks in the Lupus sample. An alternative possibility for the lack of substructure could be that the gas surface densities are high enough to drag the millimeter-emitting dust grains along such that there is little radial drift.
If this were the case, the total disk masses would be $\sim 10-20$\% of the stellar mass, near the limit of gravitational stability \citep{2024ApJ...976...50W} and the the mass of millimeter emitting dust grains is a small fraction, $\sim 5-10$\%, of the total solid mass.
Testing this scenario requires more sensitive CO observations to measure the gas disk radius.

For compact disks specifically, \citet{2025arXiv250204452T} found that such disks can form in MRI-active regions beyond the dead zone. In these regions, and particularly below $\sim$6 au, dust traps are ineffective, and the optical depth of the dust emission obscures disk substructures.

\subsection{Size-Luminosity Relation (SLR)}

The millimeter luminosity in Lupus, that increases with radius as $R_{68}^{1/\beta}$ where $\beta$ = 0.61+/-0.06, is consistent with the findings of \citet{2018ApJ...865..157A} and \citet{2017ApJ...845...44T} for a compilation of disks, as well as with \citet{2020ApJ...895..126H} for Lupus. These results align with the drift-dominated scenario described in \citet{2019MNRAS.486L..63R}. This also agrees with the super-Earth formation scenario from \citet{2024A&A...689A.236S} which requires significant radial drift. There is no clear distinction in the SLR for disks with or without observed substructures, as their radii and millimeter fluxes appear scattered. When fitted
independently, we found that the slope of disks with substructures ($\beta$ = 0.47+/-0.09), is the same as for disks without substructures ($\beta$ = 0.55+/-0.08) within error bars, and the latter group is not favored more than the other for the expected drift-dominated slope.

While our slope aligns with the drift-dominated scenario, there are still some outliers that fall below the drift-dominated slope, which we didn't address in section~\ref{sec:results}. \citet{2019MNRAS.486L..63R} predicted a population of fragmentation-dominated disks (lying closer to the fragmentation-dominated slope), which could explain some of the disks outliers. However, various factors may also contribute to these deviations in the SLR \citep{2022A&A...661A..66Z}. For instance, dust properties, such as variations in opacity or porosity, and changes to the turbulence parameter, $\alpha$, can shift a disk's position along the SLR and potentially account for some of the observed outliers. 

Specifically, higher $\alpha$ values increase the luminosity, explaining disks positioned higher on the SLR. Variations in opacity not only affect luminosity but also influence disk size as they evolve. Lastly, extremely high porosity can lead to disks with low luminosity while causing minimal changes in disk size, which could explain the disks located at the lower end of the SLR.

More recently, in Taurus, \citet{2019ApJ...882...49L} and \citet{2024ApJ...966...59S} also found an SLR with a slope consistent with drift-dominated disks by adding more observations at high-resolution, nevertheless, they also found two disks, below 5 au, that fall below this slope.

\subsection{Dust masses and optical depths}

We create a grid of radiative transfer models that predict the flux for a range of disk radii, dust masses, and stellar luminosities. Comparing our observations of size and flux, we then estimate disk masses, finding a range from 0.3 to 26.3 M$_{\oplus}$. The average optical depth for these disks shows a narrower variation and clusters around 0.7 for most disks. Most of the masses are consistently higher than those from the optically thin approximation based on flux, but they remain relatively comparable and low. These new masses remain too biased to observe any trends or changes in disk relations, such as M$_{dust}$ - R$_{dust}$ and M$_{dust}$ - M$_{*}$. When considering all disks, these relations remain unchanged. Additionally, comparing the optical depths of these disks with other parameters does not reveal any clear trends. This likely indicates that relations like M$_{dust}$ - R$_{dust}$ and M$_{dust}$ - M$_{*}$ are not significantly impacted by associated optical depth effects. While we might have expected many compact disks to be optically thick, our findings indicate otherwise. A similar pattern was observed in the rings of the DSHARP survey \citep{2018ApJ...869L..46D}, where optical depths clustered between 0.2 and 0.5, rather than being optically thick. \citet{2019ApJ...884L...5S} showed that if the dust density is regulated by planetesimal formation (small grains turning into large objects), typical millimeter continuum optical depths are $\sim 0.5$.
This means that the dust mass distribution in Lupus is likely not representative for the solid mass budget for planet formation, indicating that the bulk of the pebbles have already been sequestered in larger bodies, perhaps even planets, especially when considering efficient pebble accretion.

\subsection{Impact of compact disks in the exoplanet population}

Under the assumption that the observed rings and gaps are produced by planets, we estimate their masses using DBNets. We are in agreement with the planet masses found in previous studies, including \citet{2023ApJ...952..108Z}, \citet{2024arXiv241003823H}, \citet{2024A&A...685A..65R}, and \citet{2021A&A...645A.139K}. However, we extend their findings to include smaller substructures, with gaps as small as 4 au. As previously mentioned, the planet masses and semi-major axes for the smallest disks in our sample align closely with those of exoplanets primarily detected through radial velocity methods and where the occurrence rates of giant exoplanets peak between 1 and 10 au (\citet{2019ApJ...874...81F}; \citet{2021ApJS..255...14F}). This strengthens the case for planets forming within the observed gaps in several protoplanetary disks. Furthermore, the overlap between these gap radii and the peak in the exoplanet occurrence suggests that at least some exoplanets may form in situ in compact disks rather than migrating to their observed location or that they have already migrated to at the time the compact disks are observed.

In Fig.~\ref{fig. Dbnets-planets}, we observe an intriguing lack of planets/gap centers in Lupus disks between 5–15 au, this gap remains between 5–8 au even with additional data. Although this needs to be properly quantified, since there is additional uncertainty in the case of cavities because the planet’s location can be any value within the cavity and we need more observations of disks at high angular resolution to conclude that this "valley" is significant, we can still speculate about the possible causes if this gap is real. For this to occur, a break in the disk properties is necessary in these two regions. While there is no straightforward explanation at the moment, if the snowlines (H$_{2}$O and CO) in Lupus are located around $\sim$5 au and $\sim$15 au, this could give rise to a bimodal population of planets and gaps if planet formation is enhanced at these snowlines.

Alternatively, though less likely, we could be observing a different regime where the smallest gaps in our sample are caused by internal photoevaporation rather than planet formation \citep{2012MNRAS.426L..96O}. Disks with cavities under 10 au, such as Sz108B, J16092697-3836269, Sz72, Sz90, Sz96, Sz131, and Sz76, have accretion rates of log$_{10}$[M$_{\odot}\cdot$ yr$^{-1}]$ = -9.5, -8.1, -8.6, -8.9, -9.3, -9.1, and -9.1, respectively. Most of these align with the photoevaporation models by \citet{2018MNRAS.473L..64E} and \citet{2019MNRAS.487..691P} and could potentially be explained by this mechanism. However, J16092697-3836269 and Sz72, with larger accretion rates, fall outside the regime predicted by current photoevaporation models, suggesting that other mechanisms, like planets, may be causing these substructures. Generally it is difficult to prove whether cavities are caused by planets, photoevaporation, or a combination of the two \citep{2023A&A...679A..15G} without direct observations or limits on the planets itself. Finally, other physical mechanisms, such as dead zones, can also explain the formation of small cavities without invoking planets (\citet{2015A&A...574A..68F}; \citet{2016A&A...596A..81P} and \citet{2021A&A...655A..18G}).

For the remaining planets predicted by DBNets, Fig.~\ref{fig. Dbnets-planets} shows that they occupy a region where only a few exoplanets have been discovered (between 14 - 117 au and 20 - 2008 M$_{\oplus}$). This could be explained by three possibilities: (1) it is challenging to detect exoplanets in this region due to observation limitations that come with low planet masses or large distances from the star; (2) planets may have formed in these regions but have since migrated inward, closer to their stars, where we now observe them (\citet{2012ARA&A..50..211K}; \citet{2019MNRAS.486..453L}); or (3) the gaps observed in protoplanetary disks may not be the result of planet-disk interactions, indicating other formation processes at work (e.g snowlines \citep{2015IAUGA..2256118Z} or zonal flows \citep{2009ApJ...697.1269J}).

In Fig. \ref{fig.massvsmassgaps}, there appears to be a relation between gap widths and the ring peak positions beyond each gap, with gaps and cavities following distinct trends. Additionally, this figure also reveals a lack of substructures (ring peaks) between $\sim$13–27 AU, which is closely related to the cavity gap mentioned before in Fig.~\ref{fig. Dbnets-planets}. The absence of substructures, either cavities or rings, in this region of the disks appears to persist even when considering a more direct observable. There does not appear to be a straightforward relationship between the disk dust mass and the mass of the planet carving the gap. However, the derived planet masses are much larger than the dust masses, even though above 10 M$_{\oplus}$ most of the mass is probably in the gas envelope, this could imply that the millimeter emission we observe is coming from leftover dust particles that haven't turned into planetesimals yet, meaning that the Lupus disks could have already finished most of their planet formation. On the other hand, the observed millimeter dust emission could be just a tracer of an unseen, larger population of solids.

Assuming that for other star-forming regions compact disks around M-dwarfs are also common (\citet{2016ApJ...828...46A}; \citet{2018AJ....156...24M}),  understanding the evolutionary paths of these types of disks is crucial for planet formation and dust evolution.  The fact that we observe numerous compact disks in the Lupus region, both with and without substructures, aligns with the scenario from \citet{2021AJ....162...28V}, in which such disks, undergoing significant radial drift, can supply enough dust material to form multiple super-Earths \citep{2024A&A...689A.236S}. This offers a potential explanation for the origin of the exoplanet populations observed around M-stars to this day.
In addition, the fact that we are still observing these disks may require something to halt dust drift and trapping particles. Instead of forming a single Jupiter or Saturn-mass planet, several super-Earths or smaller planets may be forming in the inner regions of these disks, collectively stopping the drift \citep{2024arXiv241002856H}.

\section{Summary and conclusions}
\label{sec:summary}
We have analyzed a complete sample of protoplanetary disks in Lupus using the highest resolution and sensitivity data to date. The key findings of this work are as follows:
\begin{itemize}
  \item We gathered high-resolution images of Lupus disks and used new observations of 33 faint disks to complete the Lupus sample of Class II disks at high resolution. Our findings reveal that over 67\% of the disks in Lupus have dust radii smaller than approximately 30 au. Additionally, we discovered 11 new disk cavities, including one of the smallest cavities measured to date, with a radius of 4.1 au.
  \item Through visibility modeling, we measured the dust disk radii of several compact disks, finding sizes as small as 0.6 au.
  \item The observed properties of the compact disks are consistent with planet formation models which predict efficient formation of close-in Super-Earths in drift-dominated disks around M-dwarfs \citep{2024A&A...689A.236S}.
  \item We revisited the size-luminosity relation (SLR), finding good agreement with works measuring $F_{mm} \propto R_{eff}^{2}$ consistent with drift-dominated disks and extending it by including the smallest disk sizes.
  \item Comparing the total flux and radius for each disk with radiative transfer models tailored to the stellar luminosity of each source, we estimated the dust mass and average optical depths of the compact disks. Both are generally low with median values $\sim$ 3.38\,M$_\oplus$ and $\sim 0.7$ respectively. This implies that the observed dust masses in Lupus are not representative for the total solid mass budget for exoplanet formation and that the bulk of the pebbles may have already been converted into boulders or even planets or have already drifted inwards and sublimated close to the host star.
  \item We estimated planet masses for each substructure in our Lupus sample using DBNets. The smaller cavities in our study align well with the population of radial velocity detected exoplanets, while the larger ones are consistent with previous findings and they lie in a region of the parameter space where no exoplanets are currently detected. However, no clear correlation was observed between planet masses and disk dust masses in our sample.

\end{itemize}

\begin{acknowledgements}
We thank the anonymous referee for the insightful comments and suggestions, which helped improve the quality of this work.
This paper makes use of the following ALMA data: ADS/JAO.ALMA$\#$2022.1.00154.S (PI: van der Marel, Nienke), ADS/JAO.ALMA$\#$2018.1.01458.S (PI: Yen, Hsi-Wei), ADS/JAO.ALMA$\#$2017.1.00388.S (PI: Liu, Hauyu
Baobab), ADS/JAO.ALMA$\#$2018.1.00689.S (PI: Muto, Takayuki) and ADS/JAO.ALMA$\#$2022.1.01302.S (PI: Mulders,
Gijs). ALMA is a partnership of ESO (representing its member states), NSF (USA) and NINS (Japan), together with NRC (Canada), MOST and ASIAA (Taiwan), and KASI (Republic of Korea), in cooperation with the Republic of Chile. The Joint ALMA Observatory is operated by ESO, AUI/NRAO and NAOJ. This research has made use of the NASA Exoplanet Archive, which is operated by the California Institute of Technology, under contract with the National Aeronautics and Space Administration under the Exoplanet Exploration Program.
G.D.M. acknowledges support from FONDECYT project 11221206 and the ANID BASAL project FB210003.

\end{acknowledgements}

\bibliographystyle{aa}
\bibliography{bibliography}
\onecolumn
\begin{appendix}

\section{Stellar parameters Lupus}
\label{appendix:a}
\begin{table*}[!h]
\centering
\caption{Stellar Parameters of the compact disks in Lupus}
 \setlength\tabcolsep{3pt}
\renewcommand{\arraystretch}{1.2}
\centering
\begin{tabular}{cccccccc}
\hline \hline
Source& Spectral Type & T$_{\rm eff}$ [K] & L$_{*}$ [L$_{\odot}$] & M$_{*}$ [M$_{\odot}$] & log (L$_{\rm acc}$) [L$_{\odot}$] & log (M$_{\rm acc}$) [M$_{\odot}$ yr$^{-1}$] & dist [pc] \\

\hline
J16124373-3815031 & M1 & 3720 & 0.39 $\pm$ 0.27 & 0.47 & -2.1 & -9.0 & 159.85 \\
Sz117 & M3.5 & 3300 & 0.27 $\pm$ 0.19 & 0.23 & -2.3 & -8.8 & 156.95 \\
Sz110 & M4 & 3190 & 0.17 $\pm$ 0.13 & 0.18 & -2.2 & -8.7& 157.49 \\
J16134410-3736462 & M5 & 2980 & 0.03 $\pm$ 0.03 & 0.09 & -2.4 & -9.0 & 158.55 \\
J16080017-3902595 & M5.5 & 2920 & 0.04 $\pm$ 0.03 & 0.07 & -3.8 & -10.2 & 161.13 \\
Sz69 & M4.5 & 3085 & 0.08 $\pm$ 0.14 & 0.15 & -2.7 & -9.3 & 152.56 \\
Sz95 & M3 & 3410 & 0.26 $\pm$ 0.18 & 0.29 & -2.7 & -9.3 & 160.47 \\
J16085373-3914367 & M5.5 & 2920 & 0.003 $\pm$ 0.003 & 0.068 & -3.7 & -10.8 & 148.72 \\
Sz88A & M0 & 3900 & 0.3 $\pm$ 0.23 & 0.65 & -1.4 & -8.5 & 157.64 \\
J16073773-3921388 & M5.5 & 2920 & 0.01 $\pm$ 0.01 & 0.67 & -3.6 & -10.1 & 162.45 \\
J16002612-4153553 & M5.5 & 2920 & 0.07 $\pm$ 0.04 & 0.10 & -3.2 & -9.6 & 163.17 \\
Sz102 & K2 & 4710 & 0.009 $\pm$ 0.01 & - & -2.2 & - & 158.5 \\
Sz113 & M4.5 & 3085 & 0.03 $\pm$ 0.03 & 0.13 & -2.2 & -8.9 & 160.53 \\
Sz97 & M4 & 3190 & 0.10 $\pm$ 0.08 & 0.19 & -3.1 & -9.7 & 157.34 \\
J16085324-3914401 & M3 & 3410 & 0.19 $\pm$ 0.15 & 0.3 & -3.2 & -10 & 163 \\
Sz77 & K7 & 4020 & 0.59 $\pm$ 0.24 & 0.67 & -1.6 & -8.7 & 155.25 \\
Sz130 & M2 & 3560 & 0.17 $\pm$ 0.07 & 0.4 & -2.1 & -9.1 & 159.18 \\
Sz106 & M0.5 & 3810 & 0.05 $\pm$ 0.04 & 0.55 & -2.6 & -10.1 & 158.71 \\
V1192Sco & M4.5 & 3197 & 0.002 $\pm$ 0.001 & 0.17 & -4.3 & -11.8 & 147.10 \\
Sz81A & M4.5 & 3085 & 0.24 $\pm$ 0.11 & 0.18 & -2.4 & -8.8 & 158.23 \\
Sz81B & M5.5 & 3060 & 0.11 $\pm$ 0.06 & 0.13 & -3.2 & -9.6 & 158.23 \\
Sz74 & M3.5 & 3300 & 1.15 $\pm$ 0.48 & 0.3 & -1.4 & -7.8 & 158.5 \\
V856Sco & -- & -- & --  & -- & -- & -- & -- \\
J15450887-3417333 & M5.5 & 2920 & 0.06 $\pm$ 0.03 & 0.09 & -1.7 & -8.1 & 154.81 \\
J16075475-3915446 & -- & -- & --  & -- & -- & -- & -- \\
J16084940-3905393 & M4 & 3190 & 0.15 $\pm$ 0.11 & 0.19 & -3.1 & -9.6 & 160.19 \\
J15592523-4235066 & M5 & 2980 & 0.02 $\pm$ 0.01 & 0.08 & -4.4 & -11 & 147.25 \\
Sz108B & M5 & 2980 & 0.1 $\pm$ 0.08 & 0.12 & -3.0 & -9.5 & 161.22 \\
J16092697-3836269 & M4.5 & 3085 & 0.07 $\pm$ 0.05 & 0.15 & -1.5 & -8.1 & 159.19 \\
Sz72 & M2 & 3560 & 0.27 $\pm$ 0.12 & 0.37 & -1.7 & -8.6 & 156.71 \\
Sz90 & K7 & 4020 & 0.42 $\pm$ 0.28 & 0.73 & -1.8 & -8.9 & 160.37 \\
Sz96 & M1 & 3720 & 0.41 $\pm$ 0.32 & 0.46 & -2.5 & -9.3 & 155.98 \\
Sz123A & M1 & 3720 & 0.13 $\pm$ 0.09 & 0.55 & -2 & -9.1 & 162.19 \\
Sz100 & M5.5 & 2920 & 0.10 $\pm$ 0.07 & 0.13 & -3.3 & -9.6 & 158.5 \\
Sz131 & M3 & 3410 & 0.15 $\pm$ 0.06 & 0.30 & -2.3 & -9.1 & 160.62 \\
Sz73 & K7 & 4020 & 0.46 $\pm$ 0.2 & 0.7 & -0.9 & -8.0 & 157.82 \\
Sz66 & M3 & 3410 & 0.21 $\pm$ 0.09 & 0.2 & -1.7 & -8.5 & 155.92 \\
Sz65 & K7 & 4020 & 0.86 $\pm$ 0.3 & 0.6 & -2.5 & <-9.4 & 153.47 \\
Sz76 & M4 & 3190 & 0.17 $\pm$ 0.07 & 0.18 & -2.55 & -9.1 & 156.4 \\
Sz103 & M4 & 3190 & 0.11 $\pm$ 0.09 & 0.19 & -2.6 & -9.2 & 157.15 \\
Sz112 & M5 & 2980 & 0.11 $\pm$ 0.09 & 0.13 & -3.4 & -9.8 & 159.3 \\
Sz104 & M5 & 2980 & 0.06 $\pm$ 0.05 & 0.10 & -3.3 & -9.8 & 159.81 \\
J16081497-3857145 & M5.5 & 2920 & 0.009 $\pm$ 0.005 & 0.06 & -3.6 & -10.2 & 150.83 \\
HTLup & K2 & 4710 & 5.69 $\pm$ 2.1 & 1.3 & -1.1 & <-8.1 & 158.5 \\
GQLup & K6 & 4115 & 1.60 $\pm$ 0.6 & 0.6 & -0.6 & -7.4 & 154.1 \\

\hline
\label{table:Stellar parameters}
\end{tabular}

\end{table*}
\begin{table*}[!h]
\centering
\caption{Stellar Parameters of the large disks in Lupus}
 \setlength\tabcolsep{3pt}
 \renewcommand{\arraystretch}{1.3}
\centering
\begin{tabular}{cccccccc}
\hline \hline
Source& Spectral Type & T$_{\rm eff}$ [K] & L$_{*}$ [L$_{\odot}$] & M$_{*}$ [M$_{\odot}$] & log (L$_{\rm acc}$) [L$_{\odot}$] & log (M$_{\rm acc}$) [M$_{\odot}$ yr$^{-1}$]  & dist [pc] \\
\hline

J16083070-3828268 & K2 & 4710 & 1.8 $\pm$ 1.3 & 1.2 & -2.0 & <-9.2 & 158.5 \\
RYLup & K2 & 4710 & 1.84 $\pm$ 0.71 & 1.27 & -0.8 & -8.0 & 158.5 \\
Sz98 & K7 & 4020 & 1.53 $\pm$ 1.08 & 0.5 & -0.7 & -7.4 & 156.27 \\
Sz91 & M1 & 3720 & 0.2 $\pm$ 0.14 & 0.5 & -2 & -9.0 & 159.39 \\
J16090141-3925119 & M4 & 3190 & 0.09 $\pm$ 0.07 & 0.19 & -3.1 & -9.8 & 159.2 \\
Sz111 & M1 & 3720 & 0.21 $\pm$ 0.15 & 0.5 & -2.4 & -9.4 & 158.37 \\
Sz129 & K7 & 4020 & 0.42 $\pm$ 0.16 & 0.73 & -1.1 & -8.2 & 160.13 \\
RULup & K7 & 4020 & 1.46 $\pm$ 0.60 & 0.55 & -0.2 & -7 & 158.5 \\
IMLup & K5 & 4210 & 2.51 $\pm$ 1.04 & 0.72 & -1 & -7.8 & 155.82 \\
Sz114 & M4.8 & 3022 & 0.19 $\pm$ 0.14 & 0.16 & -2.6 & -9.1 & 156.76 \\
GWLup & M1.5 & 3640 & 0.32 $\pm$ 0.14 & 0.414 & -2.1 & -9.0 & 155.2 \\
Sz84 & M5 & 2980 & 0.13 $\pm$ 0.06 & 0.15 & -2.6 & -9.0 & 158.5 \\
Sz133 & K5 & 4210 & 0.07$\pm$ 0.03 & - & -1.7 & -- & 158.5 \\
Sz118 & K5 & 4210 & 0.69 $\pm$ 0.47 & 0.83 & -1.9 & -9.1 & 161.46 \\
V1094 Sco & K6 & 4115 & 1.2 $\pm$ 0.86 & 0.64 & -1.0 & -7.8 & 158 \\
RXJ1556.1 & M1 & 3705 & 0.26 $\pm$ 0.10 & 0.5 & -0.8 & -7.8 & 158 \\
MYLup & K0 & 4870 & 0.86 $\pm$ 0.33 & 1.19 & -0.6 & -8 & 158.5 \\
J16102955-3922144 & M4.5 & 3085 & 0.10 $\pm$ 0.07 & 0.15 & -3.38 & -9.9 & 160.44 \\
J16070854-3914075 & -- & -- & --  & -- & -- & -- & -- \\
J16070384-3911113 & M4.5 & 3085 & --  & -- & -5.4 & -- & 158.5 \\
J16011549-4152351 & -- & -- & --  & -- & -- & -- & -- \\
J16000236-4222145 & M4 & 3190 & 0.171 $\pm$ 0.07 & 0.19 & -2.9 & -9.4 & 160.39 \\
J16000060-4221567 & M4.5 & 3085 & 0.097 $\pm$ 0.04 & 0.15 & -3.0 & -9.5 & 159.43 \\
EXLup & M0 & 3900 & 0.73 $\pm$ 0.5 & 0.5 & -0.9 & -7.8 & 154.72 \\

\hline
\label{table:Stellar parameters 2}
\end{tabular}
\end{table*}

\clearpage
\section{Visibility models}
\label{appendix:b}
\FloatBarrier
\begin{figure*}[!h]
\centering
\includegraphics[width=1.1\columnwidth,trim={3.0cm 2.0cm 0cm 0.0cm}]{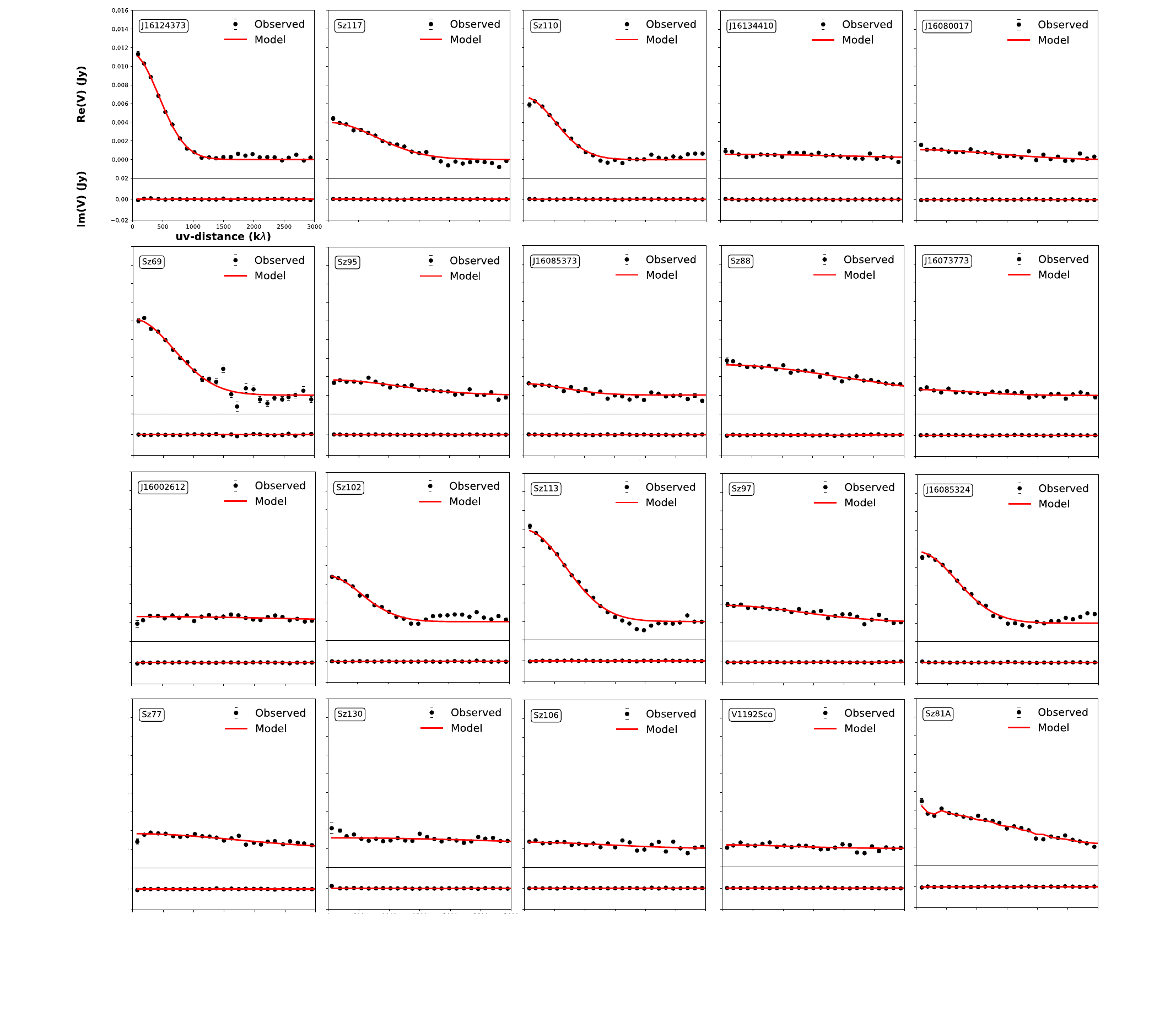}
\caption{\em{Visibility plots comparing the observed and modeled visibilities.}}
\label{fig. Visibility Models}
\end{figure*}

\begin{figure*}[!h]
\centering
\includegraphics[width=1.1\columnwidth,trim={1.0cm 1.0cm 0cm 0.0cm}]{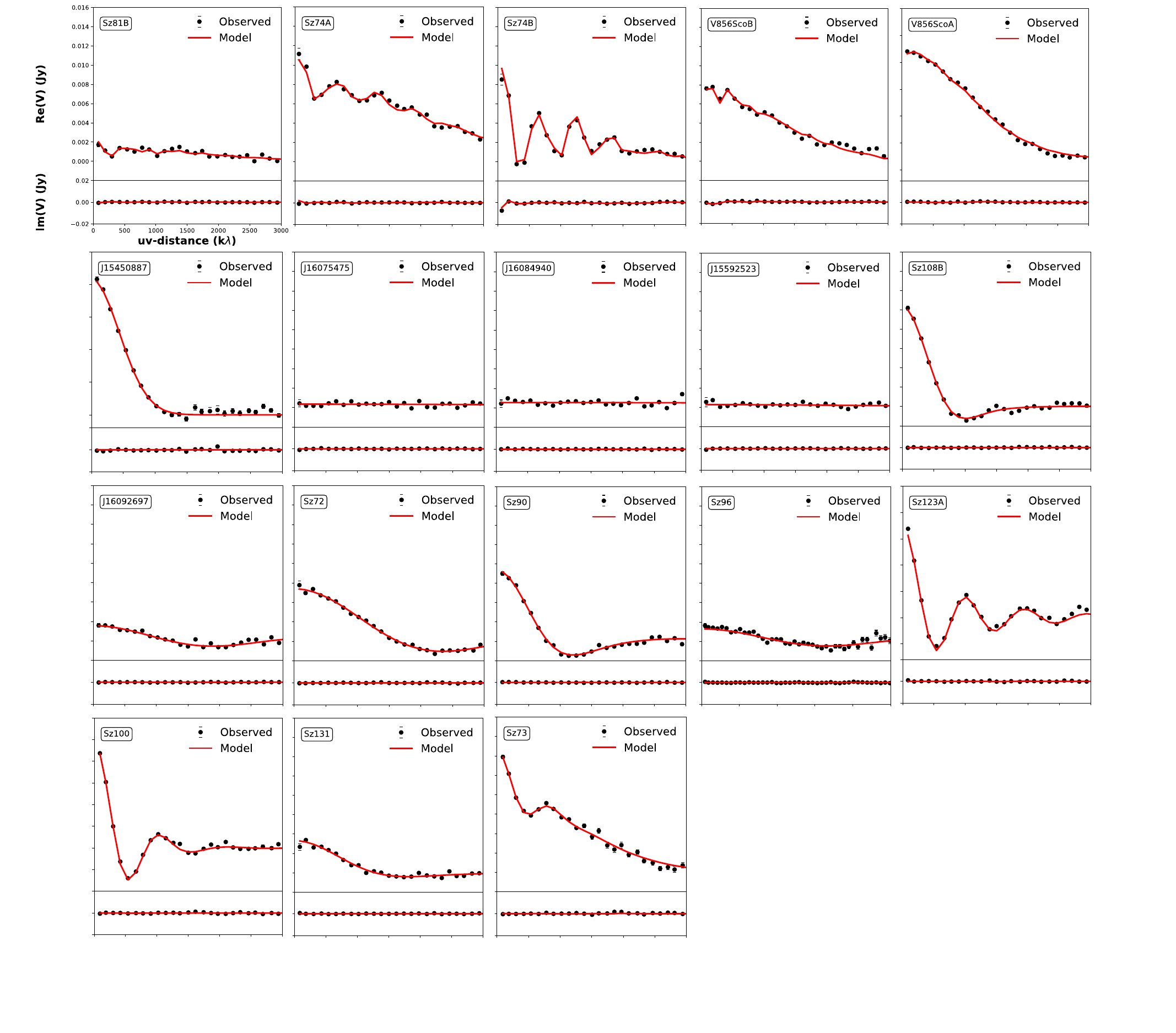}
\caption{\em{Visibility plots comparing the observed and modeled visibilities (continued).}}
\label{fig. Visibility Models(continued)}
\end{figure*}

\end{appendix}
\end{document}